\documentclass[11pt,a4wide]{article}
\usepackage{graphicx}% Include figure files
\begin{document}
\begin{center}
{\Large\bf Analytical invariant manifolds near unstable points \\
and the structure of chaos}
\vskip 0.5cm
{C. Efthymiopoulos, G. Contopoulos, and M. Katsanikas}
\vskip 0.5cm
Research Center for Astronomy and Applied Mathematics\\
Academy of Athens, Soranou Efessiou 4, 115 27 Athens, Greece
\vskip 1cm \noindent {\bf Keywords: } Hyperbolic normal form;
Invariant manifolds; Homoclinic chaos \vskip 1cm
\end{center}

\noindent{\small {\bf Abstract:} \noindent It is known that the 
asymptotic invariant manifolds around an unstable periodic orbit 
in conservative systems can be represented by convergent series
(Cherry 1926, Moser 1956, 1958, Giorgilli 2001). The unstable 
and stable manifolds intersect at an infinity of homoclinic points,
generating a complicated homoclinic tangle. In the case of simple
mappings it was found (Da Silva Ritter et al. 1987) that the domain 
of convergence of the formal series extends to infinity along the
invariant manifolds. This allows in practice to study the
homoclinic tangle using only series. However in the case of
Hamiltonian systems, or mappings with a finite analyticity domain,
the convergence of the series along the asymptotic manifolds is
also finite. Here, we provide numerical indications that the
convergence does not reach any homoclinic points. We discuss in
detail the convergence problem in various cases and we find the
degree of approximation of the analytical invariant manifolds to
the real (numerical) manifolds as i) the order of truncation of
the series increases, and ii) we use higher numerical precision 
in computing the coefficients of the series. Then we introduce a 
new method of series composition, by using action-angle variables, 
that allows the calculation of the asymptotic manifolds up to an a 
arbitrarily large extent. This is the first case of an analytic 
development that allows the computation of the invariant manifolds 
and their intersections in a Hamiltonian system for an extent long 
enough to allow the study of homoclinic chaos by analytical means.}

%%%%%%%%%%%%%%%%%%%%%%%%%%%
\section{Introduction}
%%%%%%%%%%%%%%%%%%%%%%%%%%%

It is well known that the formal series representing the invariant
manifolds (tori) near stable equilibria or stable periodic orbits
in Hamiltonian systems are not convergent in general (for a review
see Contopoulos 2002), except in the case of integrable systems.
However, near unstable equilibria, or unstable periodic orbits,
the invariant manifolds can be represented by {\it convergent}
formal series (Cherry 1926, Moser 1956, 1958) (see also  
Bruno 1971, 1989, Giorgilli 2001, Delshams and L\'{a}zaro 2005).

The motion near the unstable points takes place along such
invariant manifolds. On the other hand, near these points there
is chaos. Thus the question is how is it possible for convergent
(analytic) series to represent chaotic motions.

The structure of chaos near the unstable points is determined by
the intricate forms of the stable and unstable asymptotic
manifolds, which intersect in a complicated way forming the
so-called homoclinic or heteroclinic tangles. So far, these
structures have been studied in concrete examples mainly by
numerical means (e.g. Bertlett 1978, Bartlett 1982, Bartlett 1989,
Contopoulos 1990, Roeder et al. 2003, Contopoulos and Polymilis 
1993, Bazzani et al. 1993, Contopoulos et al. 1994, Contopoulos 
et al. 1996, Rom-Kedar 1990, Evans et al. 2004, Polymilis et al. 
2003). For computations related to celestial mechanics see Simo 
1990, Jorba and Masdemont 1999, Perozzi and Ferraz-Mello 2010, 
G\'{o}mez and Barrad\'{e}s 2011 and references therein.

Nevertheless, little has been done {\it analytically} to explore the
convergent series expansions in order to study the chaotic structures
formed by the invariant manifolds. This was due to the fact that
the domain of convergence established in the original theorems of
Moser is finite, and in general it is considered to be rather
small.

However, the question of the true size of the domain of
convergence is still open. In an early calculation, Franceschini
and Russo ( Franceschini and Russo 1981) computed several lobes 
and homoclinic points of the asymptotic manifolds in the 2D H\'{e}non 
mapping using a variant of the normal form series approach. In a more
general context, Da Silva Ritter et al. (Da Silva Ritter et al. 1987) 
have shown that in the case of some simple 2D symplectic mappings 
analytic and with an analytic inverse over the whole plane, the
domain of convergence along the asymptotic invariant manifolds
extends to {\it infinity}. (As pointed out in the historical
notes of Cabr\'{e} et al. (2005), this property is a direct 
consequence of a theorem of functional analysis going back
to Poincar\'{e} (Poincar\'{e} 1890). This property allows to
reproduce many non-trivial features of the homoclinic tangle 
using only series.

On the other hand, it was up to now unknown whether a similar
extension applies in the case of Hamiltonian flows as well (see
Vieira and Ozorio de Almeida 1996, Ozorio de Almeida and Vieira 1997).

In the present paper we make a detailed numerical study of the
size of the domain of convergence of the hyperbolic normal form
series both in 2D mappings and in 2D Hamiltonian systems. In
particular, we perform an analytical computation of the invariant
manifolds up to high truncation orders, and provide evidence
regarding i) how well they can represent the true form of the
invariant manifolds, and ii) what is the extent of the domain of
convergence.

Regarding this last question, we note, first, the well known fact
that the non-convergence or convergence of the formal series is
related to the appearance or non-appearance, in the series, of
{\it small divisors}. In the case of systems of two degrees of
freedom, near a stable periodic orbit we have in general divisors
of the form $m_1\omega_1+m_2\omega_2$, with $m_1,m_2$ integer and
$\omega_1,\omega_2$ real. Such divisors can become very small when
$|m|=|m_1|+|m_2|$ becomes large, or zero when $\omega_1, \omega_2$
are commensurable. The latter (resonant) case requires a special
treatment, which, however, presents similar non-convergence
features as the non-resonant case. On the other hand, in the case
of unstable periodic orbits we have two frequencies $\omega_1$
(real) and $\omega_2=-i\nu$ (imaginary). Then, it can be shown
that the corresponding divisors never approach very close to zero,
since they are bounded by a quantity $|m|\gamma$, where $\gamma$
is a positive constant of the order of the minimum of $|\omega_1|$
or $|\nu|$ (see Giorgilli 2001).

The first proof of the convergence of the formal series was 
provided by Cherry (1926), while a more general proof was provided 
by Moser (1956, 1958) and Giorgilli (2001). The first paper of Moser 
(1956) refers to 2D area preserving maps.
In this case, the unstable fixed points correspond to unstable
periodic orbits on the surface of section of two degrees of
freedom Hamiltonian systems. On the other hand, the second
paper (Moser 1958) investigates N-degree of freedom Hamiltonian 
systems of the form $H(x,y)=H_0(x,y)+H_1(x,y)+...$, where
$x\equiv(x_1,\ldots,x_N)$, $y\equiv(y_1,\ldots,y_N)$, are
canonical positions and momenta, and $H_i$ are polynomials
in $(x,y)$ of degree $i+2$, with a quadratic part
\begin{equation}\label{hammos}
H_0 = \sum_{j=1}^N i\omega_j x_j y_j
\end{equation}
with $\omega_1/\omega_2$ non-real, and $k_1\omega_1+k_2\omega_2
\neq \omega_j$ for all integers $k_1,k_2$ and all $j>2$.
The case of unstable equilibrium in two degrees of freedom systems 
corresponds to $\omega_1$ real,and $\omega_2=-i\nu$ imaginary, while, 
for more than two degrees of freedom, the theorem applies also to 
complex unstable points (i.e. $\omega_1,\omega_2$ complex). 
Moser's theorem then establishes
the existence of special solutions of the Hamiltonian equations
of motion in which all variables $x_j,y_j$ can be expressed via
four parameters $(q,p,\xi,\eta)$, i.e. of the form
$x_j=X_j(q,p,\xi,\eta)$, $y_j=Y_j(q,p,\xi,\eta)$, such that
for the variables $(q,p,\xi,\eta)$ we have the `normal form'
time evolution given by
\begin{equation}\label{qpxiet}
q=q_0e^{i\Omega(J,c)t},~~
p=p_0e^{-i\Omega(J,c)t},~~
\xi=\xi_0e^{\Lambda(J,c)t},~~
\eta=\eta_0e^{-\Lambda(J,c)t}~~
\end{equation}
with $\Omega=\omega_1+...$, $\Lambda=\nu+...$, and $J=iq_0p_0$,
$c=\xi\eta$. The quantities $J,c$ represent  integrals of motion
of Moser's normal form, since we have $q(t)p(t)=-iJ$, $\xi(t)\eta(t)=c$
for all times $t$. In particular:  i) the values $c=J=0$ correspond to
the unstable equilibrium point, and ii) the values $J\neq 0, c=0$
correspond to unstable periodic orbits with a frequency equal to
$\Omega(J,0)$, as well as to their {\it asymptotic orbits} lying
on the invariant manifolds of the periodic orbits. Finally,
iii) the values $c\neq 0$ correspond to orbits in the neighborhood
of the unstable periodic orbits (or points), subject to locally
hyperbolic dynamics under the Eqs.(\ref{qpxiet}).

The theorem of Moser was completed in an essential way by
Giorgilli (2001), who demonstrated that the special
solutions of Moser can be recovered via a convergent {\it
canonical} transformation of the variables $(x,y)$, such that in
the new canonical variables the Hamiltonian resumes a normal form
leading to the solutions (\ref{qpxiet}). In the theorem of
Giorgilli, one still has no small divisors, and the proof of the
convergence of the normalizing tranformation in a domain
surrounding the origin follows by a proper control of the Cauchy
estimates (see Giorgilli 2002 for a review) applying to the
derivatives of various analytic functions appearing in the
normalization canonical procedure.  For a discussion of the
convergence properties of the hyperbolic normal form series
nonlinear flows see also Bruno (1971), Delshams and L\'{a}zaro (2005).

The theorems of Moser and Giorgilli guarantee the existence of
 a disc of finite radius  of convergence of the hyperbolic
normal form series around the origin. However, the true extent
 as well as the true shape of  the domain of convergence
are not yet fully understood. Da Silva Ritter et al.
(1987) demonstrated that in the case of simple conservative 
2D mappings (i.e. mappings represented by analytic functions over 
the whole phase space) the domain of convergence of the hyperbolic 
normal form series goes to infinity along the axes $\xi=0$ and 
$\eta=0$ of the new canonical variables resulting after the mapping's 
normalizing transformation. Since these axes represent the stable 
and the unstable manifolds of the unstable point, this property 
permits in practice a purely analytic computation of the invariant 
manifolds and of all features connected with homoclinic dynamics, 
using only series (and not numerical integration methods).

In order to verify this fact, we presently compute the hyperbolic
normal form in two simple 2D symplectic mappings, namely the
standard map and the H\'{e}non map. In order to estimate the
radius of absolute convergence of the hyperbolic series along
particular directions from the origin, we use D'Alembert's
criterion. We then find that the successive radii $\rho_r$ of the
D'Alembert sequence yield larger and larger values as the
truncation order $r$ increases, thus providing evidence that the
successive values $\rho_r$ tend to infinity as
$r\rightarrow\infty$. Furthermore, we find that the rate of
increase of $\rho_r$ with $r$ is different in the case of the
standard  map than in the case of the H\'{e}non map. Namely, we
have a logarithmic dependence $\rho_r\sim log(r)$ in the former,
while we find a power-law dependence $\rho_r\sim r^p$, with
$p\simeq 2$, in the latter.  Nevertheless, when
back-transforming to the original variables, in both maps we find
that the series truncated at order $r$ represent successfully the
invariant manifolds up to an extent whose length scales as
$\sim\log(r)$. 

On the other hand, in Hamiltonian systems of two degrees of
freedom, some numerical calculations (see, for example, Fig.4 of
Vieira and Ozorio de Almeida (1996), and Fig.2 of Bongini et al. 
(2001)) give the impression  that the domain of convergence of Moser's 
normal form should extend at most up to the point where the two branches 
$\xi=0$ and $\eta=0$ (stable and unstable manifolds) of the invariant 
curve $\xi\eta=c=0$, when transformed to curves in the original
canonical variables, intersect each other at  a homoclinic point. 
According to Bongini et al. (2001) , such a point constitutes a 
``singularity'' of the formal expansions that ``defines the 
applicability limit of the normal form dynamics''.

Despite these indications, Vieira and Ozorio de Almeida (1996) 
have conjectured that the domain of convergence of Moser's normal 
form may extend long enough as to include homoclinic points. In fact, 
the same authors presented numerical calculations by which it was 
apparently possible to compute a homoclinic intersection in the 
H\'{e}non-Heiles 2D Hamiltonian model which possesses a (triple) 
hyperbolic equilibrium point. However, a careful examination of 
their calculations reveals that the computation was partly based 
on numerical propagation of the orbits and not exclusively on series. 
On the other hand, a purely analytic computation, based on series 
up to the 16th degree, failed to give a precise location of even 
one homoclinic point.

In Ozorio de Almeida and Vieira (1997), the authors invoke an argument 
according to which, using repeatedly {\it analytic continuation} 
(see our section 4), it becomes possible to obtain the time evolution 
of all initial conditions at $t=0$ along the Hamiltonian flow under
the form of a symplectic mapping connecting the initial conditions
with the values of the canonical variables at a later time $t$. As
we shall see, however, the fact that this mapping has a limited
domain of analyticity introduces a crucial difference between the
Hamiltonian and the simple mapping cases, thus, not allowing for
the extension of the domain of convergence proposed by Ozorio de
Almeida and Vieira to apply in the case of Hamiltonian flows.

Our own main results in the present paper regarding the Hamiltonian
case can be summarized as follows:

1) We  compute the hyperbolic series around unstable periodic
orbits in generic Hamiltonian systems expressed in action - angle
variables, of which the polynomial Hamiltonian systems with
unstable equilibria considered by Moser are a special case. In
such systems, we provide strong numerical indications that the
domain of convergence of the hyperbolic series is finite, and that
it contains no homoclinic intersections.

2) However, we propose a new method, which materializes the main
idea of the analytic continuation technique of Ozorio de Almeida
and Vieira (1997), and allows to obtain a parametrization of the 
invariant manifolds using only series in a domain extending well 
beyond the limit of convergence of the original series. This, 
in turn, allows to compute many homoclinic intersections and the 
lobes formed by the invariant manifolds using only series.

3) In the case of systems of non-polynomial form that can be
written in action-angle variables, the main element of our method
is to use truncated Fourier series {\it not expanded in the
angles} around the unstable point. In fact, we exploit the
property of the Fourier series that their singularity with respect
to the angles is on the imaginary axis, while all computations
with the angles lying on the real axis remain convergent.

4) We show that the accuracy of computations depends (a) on the
order of truncation of the series, and (b) on the number of digits
used in calculating the coefficients of the series.

5) Furthermore, we define `invariant' and `quasi-invariant' curves 
on the surface of section corresponding to values of $c\neq 0$. For 
small values of $c$ they correspond to initial conditions that are 
mapped onto segments of the same curve up to an extent large enough 
to include several oscillations and lobes similar to those of the 
asymptotic manifolds (see detailed definitions in section 4). 
Thus, such curves characterize the structure of chaos
in the neighborhood of the unstable periodic orbit. Also, they
have self-intersections, which, as shown in Da Silva Ritter et al. 
(1987) can be exploited in order to compute high order periodic 
orbits accumulating to one or more homoclinic points. Our analytical
method facilitates the computation of such orbits. However, a
detailed study of such computations is deferred to a future work.

The paper is organized as follows. In section 2 we present our
results for the convergence of the hyperbolic normal form series
and the analytic computation of invariant manifolds in 2D
symplectic mappings. Section 3 presents our Hamiltonian model as
well as the algorithm of computation of the hyperbolic normal form
series in the Hamiltonian case. We also find analytically the
characteristic curve, i.e., the position of the basic unstable
periodic orbit as a function of our model's perturbation parameter
$\epsilon$. Then, we study numerically the domain of convergence
in the Hamiltonian model. In section 4 we introduce our extended
analytical method of computation of the invariant manifolds, and
demonstrate its use in the computation of the structure of the
homoclinic tangle and its neighborhood. In section 5 we study
the convergence properties of the hyperbolic normal form as well
as the application of our extended method in a polynomial
Hamiltonian system which reduces to the case considered by Moser
exactly. Section 6 is a summary of our basic conclusions.

%%%%%%%%%%%%%%%%%%%%%%%%%%%%%%%%%%%%%%%%%%%%%%%%%%%%%%%%%%%%%%
\section{Analytical invariant manifolds in symplectic mappings:
convergence}
%%%%%%%%%%%%%%%%%%%%%%%%%%%%%%%%%%%%%%%%%%%%%%%%%%%%%%%%%%%%%%

In order to emphasize the difference between the Hamiltonian and the
mapping cases, we will consider first the computation of the invariant
manifolds and the convergence of the hyperbolic normal form in the
case of 2D symplectic mappings. It should be noted that such mappings 
have recently found use in the study of manifold dynamics in problems 
of astrodynamics (e.g. Ross and Scheeres 2007, Grover and Ross 2009, 
Campagnola et al. 2012). Nevertheless, as shown below, the 
domain of applicability of analytical computations of invariant 
manifolds in such cases depends crucially on the extent of the 
domain of analyticity of the original mapping. 

We consider real-analytic 2D symplectic 
mappings $M: (u_1,u_2)\rightarrow (u_1',u_2')$ of the form:
\begin{eqnarray}\label{mapgen}
u_1'=\lambda_1 u_1+F_2(u_1,u_2)+F_3(u_1,u_2)+... \nonumber \\
u_2'=\lambda_2 u_2 +G_2(u_1,u_2)+G_3(u_1,u_2)+...
\end{eqnarray}
where $\lambda_1=e^{\nu}$, $\lambda_2=\frac{1}{\lambda_1}=e^{-\nu}$,
with $\nu>0$, and $F_s$, $G_s$ are polynomials of degree $s$ in the
variables $(u_1,u_2)$. A mapping of the form (\ref{mapgen}) possesses 
an unstable fixed point at the origin $(0,0)$. In order to compute
its manifolds, we look for a near-identity canonical
transformation $\Phi\equiv(\Phi_1,\Phi_2)$ defined by
\begin{eqnarray}
u_1= \Phi_1(\xi,\eta)= \xi +\Phi_{1,2}(\xi,\eta)+\Phi_{1,3}(\xi,\eta)+...
\nonumber \\
u_2= \Phi_2(\xi,\eta)= \eta +\Phi_{2,2}(\xi,\eta)+\Phi_{2,3}(\xi,\eta)+...
\end{eqnarray}
where $\Phi_{i,s}$ are polynomials of degree $s$ in the new
canonical variables $(\xi,\eta)$, so that the mapping (\ref{mapgen})
expressed in the variables $(\xi,\eta)$ takes the normal form
\begin{eqnarray}\label{nfmap}
\xi'=W_1(\xi,\eta)=\Lambda(c)\xi=(\lambda_1+w_2c+w_3c^2+\ldots)\xi \nonumber\\
\eta'=W_2(\xi,\eta)=(1/\Lambda(c))\eta=(\lambda_2+v_2c+v_3c^2...)\eta
\end{eqnarray}
where $c=\xi\eta$, and $w_s$, $v_s$ are real constants. The unstable 
and stable manifolds correspond to the value $c=0$, and they coincide 
with the axes $\eta=0$ (unstable manifold) or $\xi=0$ (stable manifold). 
Back-transforming to the original variables yields the form of the 
invariant manifolds as functions of the co-ordinate parameters $\xi$ 
and $\eta$ respectively. 

The details of how to compute the transformation $\Phi$ are given in 
da Silva Ritter et al. (1987) (in fact, they are not different from the 
classical computation of the Birkhoff normal form for 2D mappings with 
an equilibrium point at the origin). 

We implemented the above procedure in the case of the unstable
point $(0,0)$ of the standard map
\begin{eqnarray}\label{stmap}
x_1'=x_1+K\sin(x_1+x_2)  \nonumber\\
x_2'=x_1+x_2
\end{eqnarray}
for $K=1.3$, and of the H\'{e}non map
\begin{eqnarray}\label{mapozo}
x_1'=\cosh(a)x_1 + \sinh(a)x_2 -{\sqrt{2}\over 2}\sinh(a)x_1^2  \nonumber\\
x_2'=\sinh(a)x_1+ \cosh(a)x_2 -{\sqrt{2}\over 2}\cosh(a)x_1^2
\end{eqnarray}
for $a=1.43$ (as in Da Silva Ritter et al. 1987). The mappings 
(\ref{stmap}) and (\ref{mapozo}) resume the form (\ref{mapgen}) 
after a linear symplectic diagonalizing transformation.

%--------------------------------------------------------------------------
\begin{figure}
\begin{center}
\includegraphics[scale=0.8]{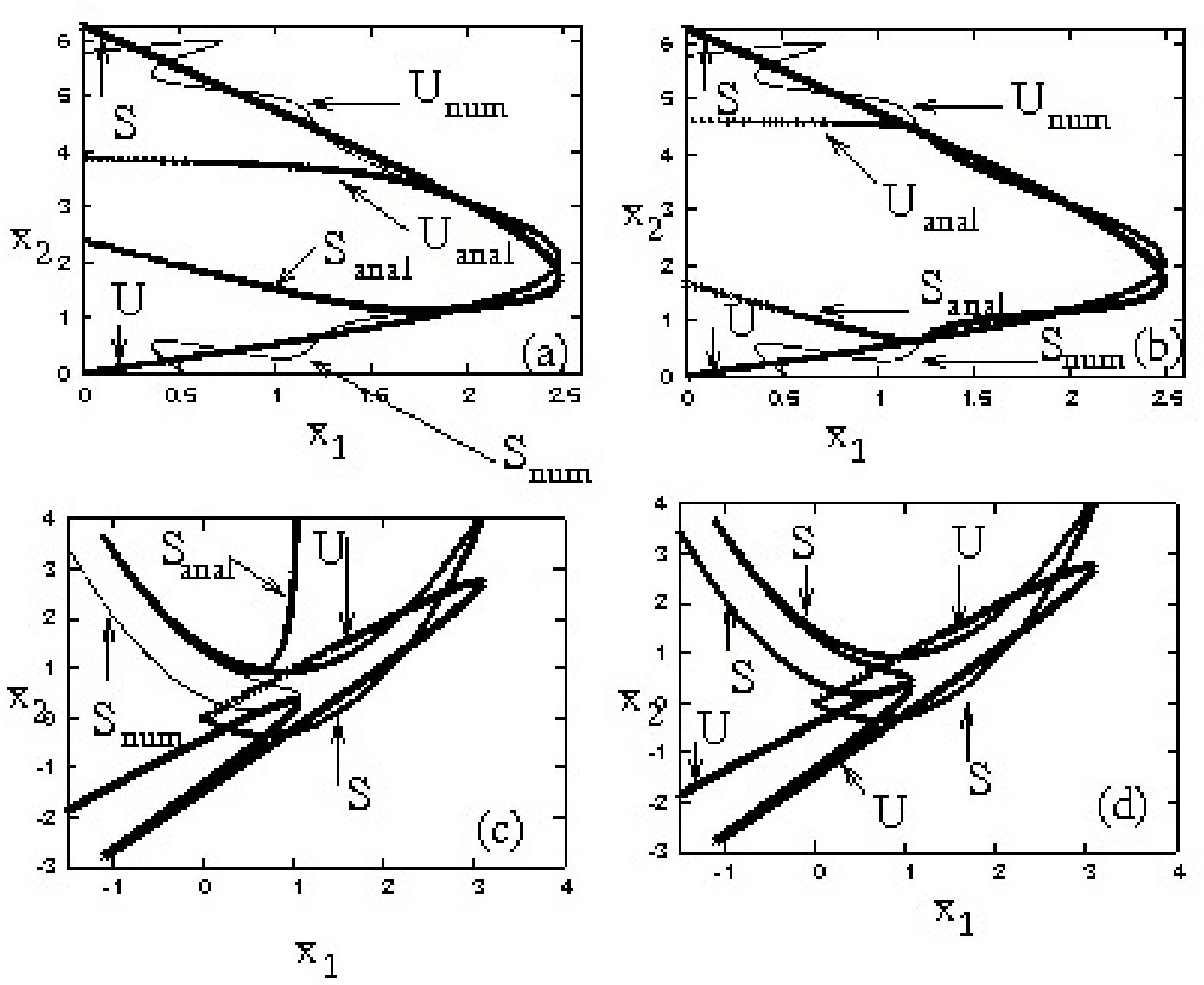}
\caption{Numerical (thin line)  and analytical calculation (thick line)
of the invariant manifolds in the case of the standard map at orders
(a) 20 and (b) 60, and in the case of the H\'{e}non map at orders
(c) 20 and (d) 60. The unstable and stable manifolds (analytical
and numerical) are denoted by U and S respectively.}
\label{mapman}
\end{center}
\end{figure}
%--------------------------------------------------------------------------
%--------------------------------------------------------------------------
\begin{figure}
\begin{center}
\includegraphics[scale=0.6]{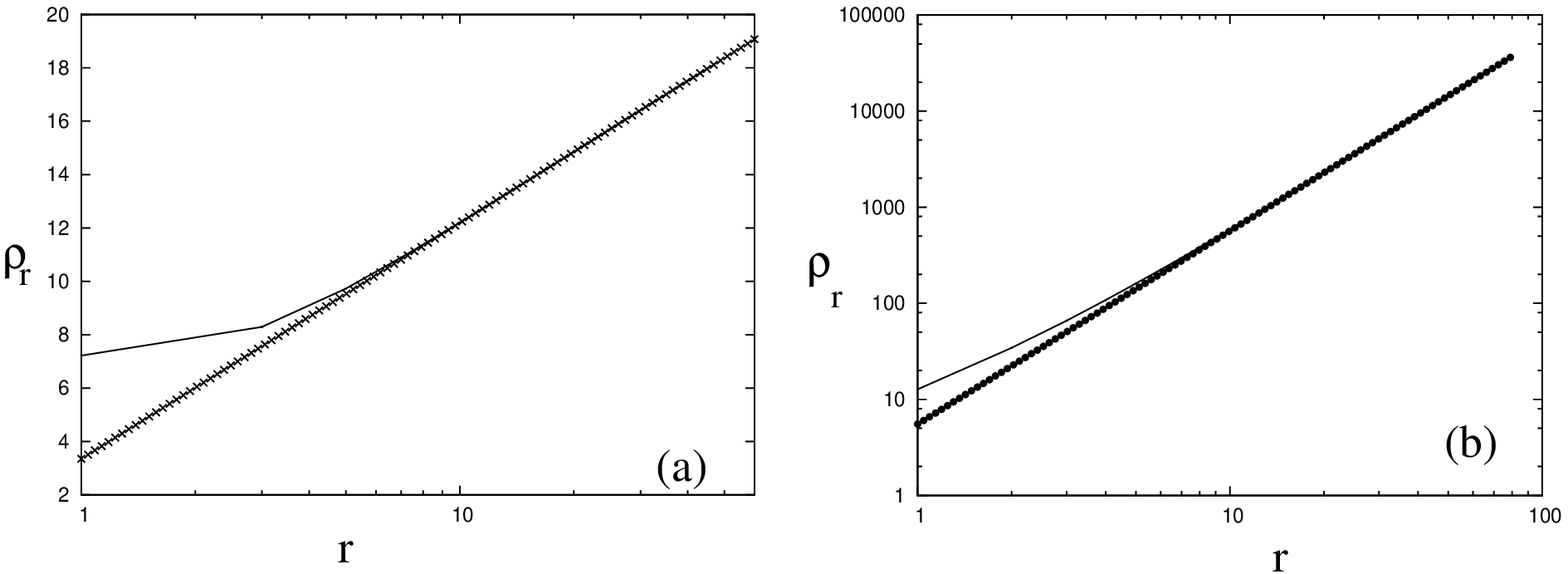}
\caption{The sequence of D'Alembert radii found by the
coefficients of the normalizing transformation $\Phi_1$ for
$\eta=0$ (see text), versus the normalization order
$r$ for (a) the standard map, and (b) the H\'{e}non map. In (a)
we find the fitting law $\rho_r= 3.34966 + 3.83952\log(r)$. In (b),
we find a power fitting law $\rho_r= 5.52 r^{2.01}$ (thick lines).}
\label{maprad}
\end{center}
\end{figure}
%--------------------------------------------------------------------------

Figures \ref{mapman}a,b,c,d show a comparison between numerical 
and analytical invariant manifolds computed by the above method, 
using the series up to the truncation orders $r=20$ (Fig.\ref{mapman}a,c), 
and $r=60$ (Fig.\ref{mapman}b,d). We note immediately that the analytical 
manifolds fit very well the numerical ones up to several homoclinic 
intersections, whose number increases with $r$. Also, the analytical
manifolds reproduce the oscillations and the lobes characteristic
of the homoclinic dynamics to an extent also increasing with $r$. 

In the case of both the above mappings, the form (\ref{mapgen}) represents 
a function analytic over the {\it whole} plane $(u_1,u_2)$. When this 
property holds true, Da Silva Ritter et al.(1987) proved that the domain 
of convergence of the normalizing transformation $\Phi$ should extend to 
infinity along the axes $\xi=0$ and $\eta=0$. In fact, we can probe 
numerically the convergence of $\Phi$ along the axes $\xi=0$ and
$\eta=0$ using {\it D'Alembert's criterion}. Namely, if we set $\eta=0$
in the normalizing transformation $\Phi$, we find expressions of the
form
\begin{eqnarray}\label{u12xi}
u_1=\xi+g_2\xi^2+g_3\xi^3+\ldots,~~~u_2=g_2'\xi^2+g_3'\xi^3+\ldots
\end{eqnarray}
The D'Alembert sequence of radii is defined by
\begin{equation}\label{rhodal}
\rho_r=\Bigg|{g_{r-1}\over g_{r}}\Bigg|,~~~~r=2,3,...
\end{equation}
The limit $\lim_{r\rightarrow\infty}\rho_r$ yields the radius of
absolute convergence along the axis $\eta=0$ (unstable manifold).
Figure \ref{maprad} shows $\rho_r$ versus $r$ for the truncation 
orders $r=2,3,...,60$ in both mappings. We have a clear indication
that $\rho_r$ goes to infinity in both cases. However, a clear 
difference is that in the case of the standard map (Fig.\ref{maprad}a) 
the increase of $\rho_r$ with $r$ seems to be logarithmic 
($\rho_r\sim\log(r)$), while in the case of the H\'{e}non map 
(Fig.\ref{maprad}b) it seems to be close to quadratic $\rho_r\sim r^2$. 
Nevertheless, even in the latter case, when we compute the length of 
the invariant manifolds in the {\it original} canonical variables up 
to the point $(\xi=\rho_r,\eta=0)$ (for the unstable manifold), we find 
that the length increases always logarithmically with $r$. This 
property sets a practical limit to the extent of the manifolds 
that can be computed analytically. At any rate, the most important 
remark is that even this limit is influenced by the extent of 
the domain of analyticity of the original mapping. Thus, for 
example, if instead of the standard mapping we use a variant of 
the form 
\begin{eqnarray}\label{stmaprat}
x_1'=x_1+{K\sin(x_1+x_2)\over 2-\cos(x_1+x_2)}  \nonumber\\
x_2'=x_1+x_2~~.
\end{eqnarray}
which possesses the same unstable point at the origin, with precisely 
the same eigenvalues and eigenvectors as the standard map (\ref{stmap}), 
we find that the analytical computation of the invariant manifolds in 
that case diverges beyond a domain that does not include any homoclinic 
point. This is due to the fact that the mapping (\ref{stmaprat}) 
has a finite domain of analyticity $|x_1+x_2|<1.31696$. Thus, 
the analytical computation of the invariant manifolds cannot go 
beyond this domain. Note that singularities of the form of the 
map (\ref{stmaprat}) are typical in applications of astrodynamics, 
where the domain of analyticity cannot extend beyond the distance 
of the considered unstable equilibrium point from any one of the 
bodies.  

We will then pass to our main question, i.e. how do the above
results compare with the extent of the domain of convergence of
the hyperbolic normal form series in the Hamiltonian case.

%%%%%%%%%%%%%%%%%%%%%%%%%%%%%%%%%%%%%%%%%%%%%%%%%%%%%%%%%%%%%%%
\section{Analytic manifolds in a Hamiltonian model}
%%%%%%%%%%%%%%%%%%%%%%%%%%%%%%%%%%%%%%%%%%%%%%%%%%%%%%%%%%%%%%%

%-----------------------
\subsection{The model}
%-----------------------
We consider first a Hamiltonian system of one and a half
degrees of freedom corresponding to a periodically driven pendulum
model with explicit dependence on time. The Hamiltonian is
\begin{equation}\label{hrpr}
H={p^2\over 2} - \omega_0^2(1+\epsilon(1+p)\cos\omega t)\cos\psi~~.
\end{equation}
Introducing a dummy action $I$, and its conjugate angle
$\phi=\omega t$, with $\omega=1$, the Hamiltonian can be written 
equivalently as
\begin{equation}\label{hrpr2}
H(\psi,\phi,p,I)={p^2\over 2}
+ \omega I-\omega_0^2(1+\epsilon(1+p)\cos\phi)\cos\psi~~.
\end{equation}

\begin{figure}
\begin{center}
\includegraphics[scale=0.8]{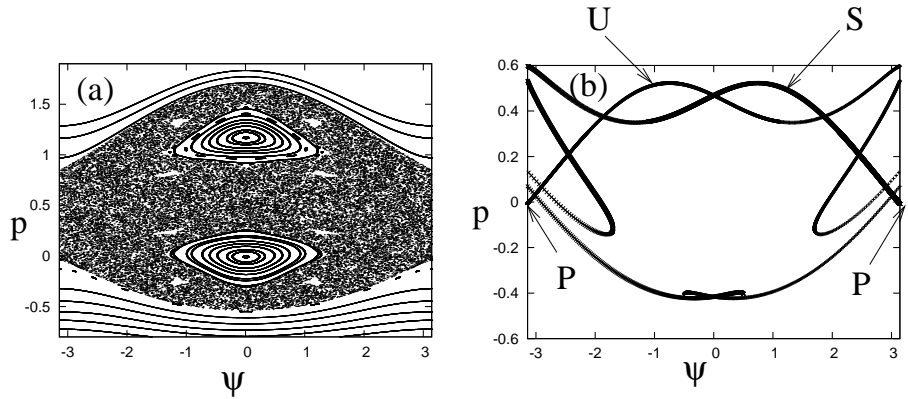}
\caption{(a) The phase portrait (stroboscopic surface of section
$(\psi(nT),p(nT))$ for the Hamiltonian system (\ref{hrpr2})
for $\epsilon=1$, $\omega_0=0.2\sqrt{2}$ and  $\omega=1$.
(b) The form of the stable ($W_S$) and unstable $W_U$ manifolds
of the main unstable periodic orbit P, given modulo $2\pi$
at the above parameter values.}
\label{phase}
\end{center}
\end{figure}

Figure \ref{phase}a shows the phase portrait for $\epsilon=1$,
when $\omega_0=0.2\sqrt{2}$, $\omega=1$. The phase portrait is
obtained by a stroboscopic plot of all points $(\psi(nT),p(nT))$
along particular orbits at the successive times $t=nT$, $n=1,2,...$,
where $T=2\pi/\omega$ is the perturber's period. We observe that
in the libration domain and in a large domain above it (with
$p>0$) most trajectories are chaotic. We distinguish only a
large island in the libration domain, as well as one conspicuous
1:1 resonant island and some other smaller islands hosting
quasi-periodic trajectories. Above and below the chaotic
domain, we have many rotational invariant KAM curves.
%-------------------------------------------------------------------
%-------------------------------------------------------------------

The principal domain of chaos is caused by the homoclinic tangle of
the invariant manifolds of a main unstable periodic orbit (denoted
hereafter as P) of the Hamiltonian (\ref{hrpr2}), which is formed
as the continuation of the unstable equilibrium point at $\psi=\pi$
(or $-\pi$), $p=0$ of the pendulum case ($\epsilon=0$).
For $\epsilon\neq 0$, this periodic orbit yields a fixed point
$\psi_0=\pm\pi$, $p_0(\epsilon)<0$ on the stroboscopic surface
of section $(\psi(nT),p(nT))$. For $\epsilon=1$, we have
$p_0=-7.32464987\times 10^{-3}$.

The intersection of the invariant manifolds of P with the surface
of section $(\psi(nT),p(nT))$ is shown in Fig.\ref{phase}b. This
is a purely numerical computation of the manifolds, obtained by
iterating an initial segment of length $DS=10^{-4}$ along the
unstable or the stable eigendirections, in the forward or backward
sense of time respectively. Due to the particular form of the
Hamiltonian (\ref{hrpr}), the manifolds, as in fact the whole
chaotic domain, have well developed lobes in the upper part of the
chaotic layer (for $p>0$), but not so much in the lower part which
is delimited by the presence of rotational KAM curves.

%----------------------------------
\subsection{Hyperbolic normal form}
%----------------------------------
In order to compute a hyperbolic normal form for the Hamiltonian
(\ref{hrpr2}), we first Taylor-expand the Hamiltonian (\ref{hrpr2})
in the neighborhood of the point
$\psi_0=\pi$ (or, equivalently, $-\pi$). Setting $\psi=\pi+u$,
up to fourth order we have
\begin{equation}\label{hamex4}
H={p^2\over 2}+\omega I
-0.08\left(1+0.5\epsilon(1+p)(e^{i\phi}+e^{-i\phi})\right)
\left(-1+{u^2\over 2}-{u^4\over 24}-...\right)~~.
\end{equation}
From this development we see that $u$ (and $\psi$)corresponds
to the hyperbolic degree of freedom, while $\phi$ corresponds
to the elliptic degree of freedom. The difference between the
two angles appearing in the hyperbolic and elliptic degrees
of freedom is crucial in the discussion of section 4 below.
The hyperbolic part of the Hamiltonian has the form
$H_h=p^2/2-\nu^2 u^2/2$ where $\nu^2=0.08$. We then introduce
the linear canonical transformation:
\begin{equation}\label{hlintra}
p={\sqrt{\nu}(\xi+\eta)\over\sqrt{2}},~~~
u={(\xi-\eta)\over\sqrt{2\nu}}~~~
\end{equation}
where $\xi$ and $\eta$ are the new canonical position and
momentum respectively. Then $H_h$ acquires the form
$H_h=\nu\xi\eta$.

For the computation of the hyperbolic normal form, we introduce a
variant of the algorithm proposed by Giorgilli (2001). The main
change is that for the oscillatory degree of freedom we keep the
action-angle form of the corresponding variables in the transformed
Hamiltonian. In this way, we avoid the need to introduce expansions
in cartesian variables for this degree of freedom. A general
Hamiltonian of this form is
\begin{equation}\label{hamgen}
H(\phi,I,\xi,\eta)=\omega I + \nu\xi\eta +
H_1(\phi,I,\xi,\eta)
\end{equation}
where i) $\nu$ is a real constant, ii) $H_1$ is periodic in
$\phi$, and admits a convergent polynomial expansion in the variables
$(\xi,\eta)$, and iii) $H_1$ is analytic in a complexified domain
$I\in\mathbf{C}$, $|I|<\rho$, $|Im(\phi)|<\sigma$, for two positive
constants $(\rho,\sigma)$. Similar results apply in models with 
more than one elliptic degrees of freedom (as, for example, 
the `bicircular' model of astrodynamics, see G\'{o}mez et al. 
2001).  

Close to the main unstable periodic orbit of the Hamiltonian
(\ref{hrpr2}), we look for a transformation from old to new
canonical  variables $\Phi: (\xi,\phi,\eta,I)$
$\rightarrow$ $(\xi',\phi',\eta',I')$, of the form
\begin{eqnarray}\label{trhyp}
\xi&=&\Phi_{\xi}(\xi',\phi',\eta',I')\nonumber\\
\phi&=&\Phi_\phi(\xi',\phi',\eta',I')\\
\eta&=&\Phi_{\eta}(\xi',\phi',\eta',I')\nonumber\\
I&=&\Phi_I(\xi',\phi',\eta',I')\nonumber
\end{eqnarray}
so that the Hamiltonian in the new variables takes the form:
\begin{equation}\label{nfhyp}
Z_h = \omega I'+ \nu \xi'\eta' + Z(I',\xi'\eta')~~.
\end{equation}
In a Hamiltonian like (\ref{nfhyp}), the quantities $I'$ and $c=\xi'\eta'$
are integrals of motion. Furthermore, for any value of $I'$, the point
$\xi'=\eta'=0$ corresponds to a periodic orbit, since, from Hamilton's
equations we find $\dot{\xi}'=\dot{\eta}'=0$, while $\phi'=\phi'_0
+(\omega+\partial Z(I',0)/\partial I')t$. This implies a periodic
orbit, with frequency $\omega'=(\omega+\partial Z(I',0)/\partial I')$.
Note that in a system like (\ref{hrpr2}), where the action $I$ is
dummy, $I'$ appears in the hyperbolic normal form only through the
term $\omega I'$. Thus, in this case the periodic solution
$\xi'=\eta'=0$ has a frequency always equal to $\omega$.

By linearizing Hamilton's equations of motion near this solution,
we find that it is always unstable. In fact, we can easily show
that the linearized equations of motion for small variations
$\delta\xi',\delta\eta'$ around $\xi'=0,\eta'=0$ are
$$
\dot{\delta\xi'}=(\nu+\nu_1(I'))\delta\xi',~~~~
\dot{\delta\eta'}=-(\nu+\nu_1(I'))\delta\eta'
$$
where $\nu_1(I')=\partial Z(I',\xi'\eta'=0)/\partial(\xi'\eta')$.
The solutions are $\delta\xi'(t)=\delta\xi_0' e^{(\nu+\nu_1)t}$,
$\delta\eta'(t)=\delta\eta_0' e^{-(\nu+\nu_1)t}$. After one period
$T=2\pi/\omega$ we have $\delta\xi'(T)=\Lambda_1\delta\xi_0'$,
$\delta\eta'(T)=\Lambda_2\delta\xi'_0$, where $\Lambda_{1,2}=e^{\pm
2\pi(\nu+\nu_1)/\omega}$. Thus, the two eigendirections of the
linearized flow correspond to setting $\delta\xi_0'=0$, or
$\delta\eta_0'=0$, i.e. they coincide with the axes $\xi'=0$, or
$\eta'=0$. These axes are invariant under the flow of (\ref{nfhyp})
and, therefore, they constitute the unstable and stable manifolds
of the associated periodic orbit P in the new variables $\xi',\eta'$.

If we specify the form of the transformation (\ref{trhyp}), we can
pass from the variables $(\xi',\eta')$ to the variables $(\xi,\eta)$,
thus finding the form of the invariant manifolds in the original
canonical variables as well.

In order to specify the transformation (\ref{trhyp}), we perform
a step by step normalization procedure, using a computer-algebraic
program to perform expansions up to a high order. To this end,
we split the original Hamiltonian (\ref{hamgen}) in terms of
different orders of smallness, by introducing an artificial
parameter $\lambda$, called the `book-keeping' parameter, with
numerical value equal to $\lambda=1$ (see Efthymiopoulos 2012a).
To every term in the Hamiltonian expansion we then introduce a
factor of the form $\lambda^r$, indicating that the term is
of r-th order of smallness. The book-keeping rule is the following:
all monomial terms in the Hamiltonian containing a product of the form
$\xi^{s_1}\eta^{s_2}$ acquire a book-keeping factor $\lambda^{s_1+s_2-2}$.
After this, we multiply by one more factor $\lambda$ all the terms
multiplied by $\epsilon$.

ii) {\it Hamiltonian normalization}.
The Hamiltonian normalization is accomplished by means of
Lie series (see Giorgilli 2001, or Efthymiopoulos 2012a) via
the following recursive algorithm: after $r$ consecutive
transformations, we pass to new canonical variables
$(\xi,\phi,\eta,I)$ $\rightarrow$ $(\xi^{(1)},\phi^{(1)},
\eta^{(1)},I^{(1)})$ $\rightarrow$ $\ldots$ $\rightarrow$
$(\xi^{(r)},\phi^{(r)},\eta^{(r)},I^{(r)})$, in which the
Hamiltonian has the form:
\begin{equation}\label{hamr}
H^{(r)}=Z_0+\lambda Z_1 + ... + \lambda^r Z_r + \lambda^{r+1}
H^{(r)}_{r+1}+ \lambda^{r+2} H^{(r)}_{r+2}+\ldots
\end{equation}
where $Z_0=\omega I^{(r)} +\nu\xi^{(r)}\eta^{(r)}$. The Hamiltonian
term $H^{(r)}_{r+1}$ contains a sum of terms that are not in normal
form, denoted by $h^{(r)}_{r+1}$. The $(r+1)-$th order Lie generating
function $\chi_{r+1}$ is the solution of the homological equation
\begin{equation}\label{homor}
\{Z_0,\chi_{r+1}\}+\lambda^{r+1}h^{(r)}_{r+1}=0~~
\end{equation}
where $\{\cdot,\cdot\}$ denotes the Poisson bracket operator.
We then compute the new transformed Hamiltonian via
\begin{equation}\label{hamrp1}
H^{(r+1)}=\exp(L_{\chi_{r+1}})H^{(r)}~~
\end{equation}
where $L_{\chi}=\{\cdot,\chi\}$.
This is in normal form up to terms of order
$r+1$, namely:
\begin{equation}\label{hamrp1n}
H^{(r+1)}=Z_0+\lambda Z_1 + ... + \lambda^r Z_r + \lambda^{r+1}
Z_{r+1}+ \lambda^{r+2} H^{(r+1)}_{r+2}+\ldots
\end{equation}
where $Z_{r+1}=H^{(r)}_{r+1}-h^{(r)}_{r+1}$.

The solution of the homological equation is found by
noting that the action of the operator $\{Z_0,\cdot\}=
\{\omega I+\nu \xi\eta,\cdot\}$ on monomials of the
form $\xi^{s_1}\eta^{s_2}a(I)e^{ik_2\phi}$ yields
$$
\left\{\omega I+\nu \xi\eta
,\xi^{s_1}\eta^{s_2}a(I)e^{ik_2\phi}\right\}
=
-[(s_1-s_2)\nu+i\omega k_2]\xi^{s_1}\eta^{s_2}a(I)e^{ik_2\phi}~~.
$$
We write $h^{(r)}_{r+1}$ as
$$
h^{(r)}_{r+1}=\sum_{(s_1,s_2,k_2)\notin{\cal M}}
b_{s_1,s_2,k_2}(I)\xi^{s_1}\eta^{s_2}e^{ik_2\phi}
$$
where we omit, for simplicity, superscripts from the notation of
all canonical variables, and we set ${\cal M}$ as the set:
\begin{equation}\label{resmodhyp}
{\cal M}=\left\{(s_1,s_2,k_2): s_1=s_2~\mbox{and}~k_2=0\right\}~~.
\end{equation}
Then
\begin{equation}\label{chi1hyp}
\chi_{r+1}=\sum_{(s_1,s_2,k_2)\notin{\cal M}} {b_{s_1,s_2,k_2}(I)\over
(s_1-s_2)\nu+i\omega k_2} \xi^{s_1}\eta^{s_2}e^{ik_2\phi}~~.
\end{equation}
After computing $\chi_{r+1}$, we compute the next normalized Hamiltonian
$H^{(r+1)}$ via Eq.(\ref{hamrp1}). We truncate all expressions up to a
given maximum order $r_{max}$. This completes one step of the normalization
algorithm.

Setting the maximum normalization order equal to the truncation order
$r_{max}$, we approximate the new canonical variables corresponding to
the final computed normal form as
$$
(\phi',\xi',\eta',I')\simeq(\xi^{(r_{max})},\phi^{(r_{max})},
\eta^{(r_{max})},I^{(r_{max})})~~.
$$
The transformation (\ref{trhyp}) is then approximated by the truncated
form of the Lie series composition:
\begin{eqnarray}\label{oldnew}
\xi&=&\exp(L_{\chi_{r_{max}}})
\exp(L_{\chi_{r_{max}-1}})\ldots\exp(L_{\chi_{1}})
\xi^{(r_{max})}\nonumber\\
\phi&=&\exp(L_{\chi_{r_{max}}})
\exp(L_{\chi_{r_{max}-1}})\ldots\exp(L_{\chi_{1}})
\phi^{(r_{max})}\\
\eta&=&\exp(L_{\chi_{r_{max}}})
\exp(L_{\chi_{r_{max}-1}})\ldots\exp(L_{\chi_{1}})
\eta^{(r_{max})}\nonumber\\
I&=&\exp(L_{\chi_{r_{max}}})
\exp(L_{\chi_{r_{max}-1}})\ldots\exp(L_{\chi_{1}})
I^{(r_{max})}\nonumber~~.
\end{eqnarray}
The computation of the transformation (\ref{oldnew}) completes the
implementation of the hyperbolic normal form algorithm. In the
computer implementation, the steps are:
i) We compute the form of all generating functions $\chi_1$, $\ldots$,
$\chi_{r_{max}}$ as well as the form of the final transformed Hamiltonian
$H^{(r_{max})}$.
ii) We compute the canonical transformation (\ref{oldnew}). In the
particular case of the Hamiltonian (\ref{hrpr2}), since the action
$I$ is a dummy variable, we have the simplifying properties that
a) $\phi=\phi^{(r)}$ for all $r>0$, and b) the dummy action $I^{(r)}$
does not appear in the transformation equations for the variables
$\xi$ and $\eta$. iii) We identify $\xi^{(r_{max})},\eta^{(r_{max})}$
as the best possible approximants to the `new' canonical variables
$\xi',\eta'$. Then, for computing various structures, we assign values
to $\eta'$ and to $\xi'$, for fixed values of $\phi'$. We then compute
the values of the original variables $(\xi,\eta)$ by the transformation
(\ref{trhyp}), and hence, through Eq.(\ref{hlintra}) the values of
$p$ and $\psi=\pi+u$.

%----------------------------------------------------------------------
\subsection{Position of the main unstable periodic orbit}
%----------------------------------------------------------------------
A property of the transformations (\ref{oldnew}) is that, setting
$\xi'=\eta'=0$ allows for a computation of the whole form of the
unstable periodic orbit corresponding to the origin, in the new
variables, for various values of the (constant) action $I'$.
This is because in the new variables, for initial conditions
$I'(0)=I_0$, $\phi'(0)=\phi_0$ and $\xi'(0)=\eta'(0)=0$, the
normal form (\ref{nfhyp}) implies the trivial evolution
$\xi'(t)=\eta'(t)=0$, and $I'(t)=I_0$, $\phi'(t)=\phi_0
+\left(\omega+\partial Z(I',0)/\partial I'|_{I'=I_0}\right)t$.
Substituting these expressions into (\ref{trhyp}), we compute
the functional form of the time evolution of all quantities
$\phi(t)$, $I(t)$, and $\xi(t),\eta(t)$ for the periodic orbit
expressed in the original variables as well. Finally, using
the linear transformation (\ref{hlintra}) we obtain the
same form in the variables $p(t)$ and $\psi(t)$. We note
that in this way we can obtain the analytic representation
of the periodic orbit in Fourier series depending periodically
on the time without using Lindstedt series (cf. Bell\'{o} 
et al. 2010).

For the Hamiltonian (\ref{hrpr2}), up to terms of second degree
in the book-keeping parameter $\lambda$, we find the following
formula for the periodic orbit P:
\begin{eqnarray}\label{per}
\psi_P(t)&=&\pi+0.0740741\epsilon\sin t
- 0.000726216\epsilon^2\sin(2t)+...\\
p_P(t)&=&-0.00592593\epsilon\cos t -
0.00145243\epsilon^2\cos(2t)+....\nonumber
\end{eqnarray}
We note that both functions $\psi_P(t)$ and $p_P(t)$ do not depend
on the action $I'$. As explained above, this is a feature of the
specific model considered.

%-------------------------------------------------------------------
\begin{figure}[t]
\begin{center}
\includegraphics[scale=0.47]{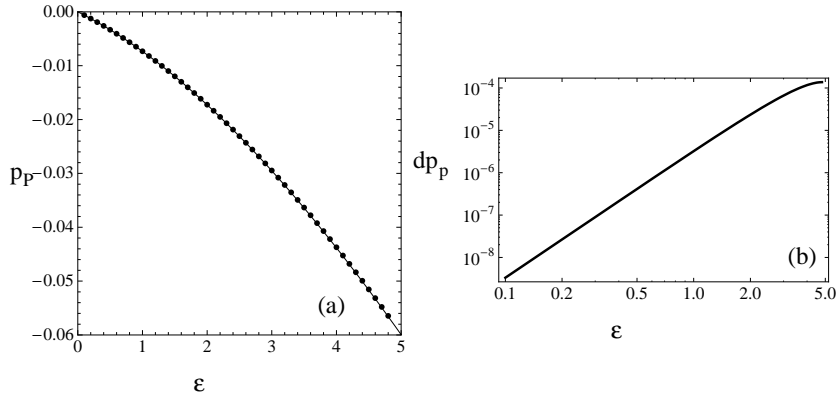}
\caption{(a) The characteristic curve (position of the initial condition
$p_P$ (for $\psi_0=\phi_0=0$) of the main unstable periodic orbit
P for the Hamiltonian system (\ref{hrpr2}) as a function of the
perturbation parameter $\epsilon$, found analytically (solid
line) and by a Newton-Raphson numerical method (points). (b) Difference 
between numerical and analytical solution at the truncation order $r=15$.}
\label{pendchar}
\end{center}
\end{figure}
%-------------------------------------------------------------------
Setting $t=0$ in Eqs.(\ref{per}), we obtain the position of the
periodic orbit on the surface of section as a function of
$\epsilon$. Figure \ref{pendchar}a shows the so-called
{\it characteristic curve} of P. In our case, since we always
have $\psi_P=0$, the characteristic curve simply consists of
plotting the initial condition of the periodic orbit $p_P(0)$
on the surface of section versus $\epsilon$. The dotted curve shows
$p_P(0)$ as a function of $\epsilon$, computed by a purely numerical
process, i.e., implementing Newton's root-finding method, while the
solid curve yields $p_P(\epsilon)$ as computed by a hyperbolic normal
form at the normalization order $r=15$. Figure \ref{pendchar}b shows 
the absolute difference between the numerical and the analytical 
determination as function of $\epsilon$. For fixed $r=15$, the 
error is about $3\times 10^{-9}$ for $\epsilon=0.1$ and rises 
up to $10^{-4}$ for the highest considered value $\epsilon=5$. 
However, since the domain of convergence of the hyperbolic normal 
form includes always the origin, the computation of the characteristic 
curve, as well as of the whole form of the periodic orbits, can be made
with arbitrarily high precision for all values of $\epsilon$ for
which we have convergence, by increasing the order of $r$. Thus, 
by raising $r$ to about $r=50$ we are able to recover eight 
significant digits for values of $\epsilon$ as high as 
$\epsilon=5$, despite the fact that the motions in the neighborhood 
of the periodic orbit are very chaotic for such a high value of 
$\epsilon$.

%-----------------------------------------------------------------
\subsection{Analytical invariant manifolds. Domain of convergence}
%-----------------------------------------------------------------
The unstable manifolds emanating from P in the case of the Hamiltonian
model (\ref{hrpr2}) are computed by setting $\eta'=0$ in the
transformation (\ref{trhyp}). The manifolds are two dimensional,
and they are given parametrically by various values of $\xi'$ for
a fixed value of the angle $\phi'$. Setting $\phi'=0$ yields the
asymptotic curves, i.e., the one-dimensional intersections of the
manifolds with the stroboscopic surface of section.

Our first computation refers to the integrable case $\epsilon=0$.
Then, the Hamiltonian (\ref{hrpr2}) becomes simply the pendulum
Hamiltonian, and the unstable and stable manifolds simply coincide
with the pendulum separatrix. Furthermore, since we are only left
with terms containing one couple of canonical variables ($\xi,\eta$),
we are able to perform high order, and high precision, computations
using Mathematica.

%----------------------------------------------------------------------
\begin{figure}
\begin{center}
\includegraphics[scale=0.85]{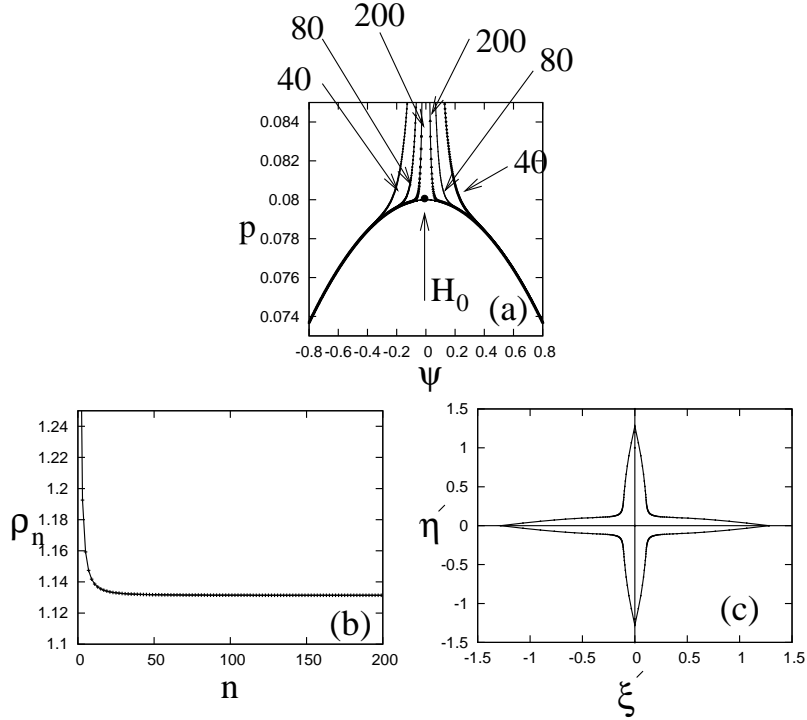}
\caption{(a) A focus on the uppermost part of the separatrix of 
the pendulum model (Eq.(\ref{hrpr2}) for $\epsilon=0$), showing 
the exact separatrix (thin curve) superposed by the analytical
computation of the unstable manifolds (left curves, deviating
upwards) and of the stable manifolds (right curves, deviating
upwards) at the normalization orders $r=40$, $r=80$, and $200$.
As $r$ increases, the analytical curves approach closer and
closer to the point in the middle of the separatrix, where
$\psi=0$. (b) The D'Alembert requence $\rho_n$ for the normalizing
transformation, setting $\eta=0$ (see text). In this case,
the radius of absolute convergence turns to be finite, as
$\rho_n$ stabilizes to $\rho_n=0.1392$ for larger $n$.
(c) Numerical estimation of the boundary of the domain of
convergence around the origin, using the values of the D'Alembert
radii computed in different directions on the plane $(\xi'\eta')$.}
\label{mansep}
\end{center}
\end{figure}
%----------------------------------------------------------------------
Figure \ref{mansep}a shows the comparison between the exact
separatrix and an analytical computation of the asymptotic
manifolds using the hyperbolic normal form at different truncation
orders of the normalizing transformation $\Phi$, namely  a)
$r_{max}=40$, b) $r_{max}=80$, and c) $r_{max}=200$, for
$\omega_0=0.2\sqrt{2}=0.2828$. We observe immediately the main
effect, namely that the analytical manifolds, even after a
truncation order as high as 200, deviate from the separatrix a
little before the point $H_0\equiv(u=\pi,p=2\omega_0)$. For
$\epsilon=0$ this is the point where the upper branch of the
unstable manifold joins smoothly with the upper branch of the
stable manifold of P, while, as we will see, for arbitrarily small
$\epsilon$, the point $H_0$ corresponds to a homoclinic
intersection of the two manifolds, which takes place always at the
angle $u=\pi$ (corresponding to $\psi=0$).

%---------------------------------------------------------------
\begin{figure}
\begin{center}
\includegraphics[scale=0.8]{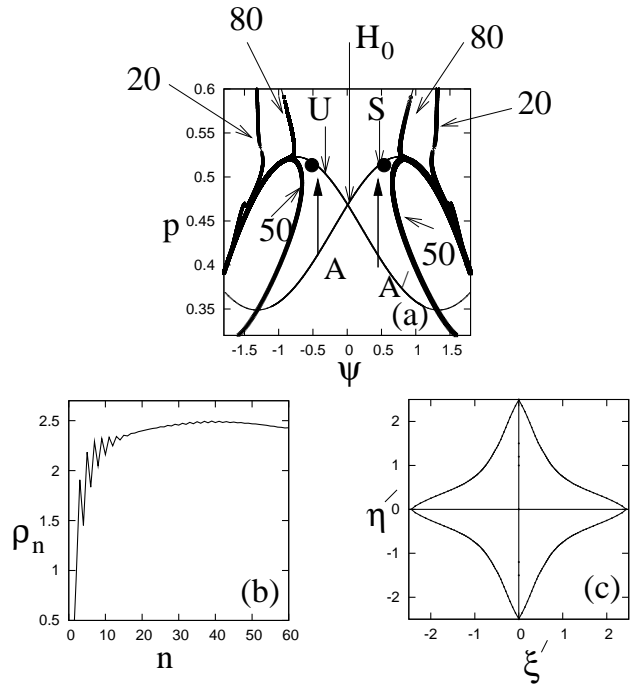}
\caption{(a) Same as in Fig.\ref{mansep}a, but for the asymptotic
manifolds of P when $\epsilon=1$. The analytical manifolds are
computed at orders $r=20$ (deviating upwards), $r=50$ (deviating
downwards), and $r=80$ (deviating upwards). We observe that no
important improvement exists between the orders 50 and 80. A and
A' are the limits of convergence of the series along the unstable
(U) and stable (S) manifolds. (b) The sequence of d'Alembert radii
also indicates a finite radius of absolute convergence along the
manifolds at $\rho\sim 2.4$. (c) The domain of absolute
convergence, found as in Fig.\ref{mansep}c.} \label{eps1man}
\end{center}
\end{figure}
%---------------------------------------------------------------
The fact that we use an 80-digit precision is necessary in order to
check the accuracy of our computations. In fact, using the usual
16-digit precision, we find that the accumulation of round-offs in
the calculation of the hyperbolic normal form coefficients results
in a number of coefficients being affected up to their first digit
after the truncation order $r=56$.

The deviation of the analytical manifolds from the true separatrix
shown in Fig.\ref{mansep}a is a phenomenon observed also in Vieira
and Ozorio de Almeida (1996; their Fig.4). In fact,
these authors' computation also points towards the fact that the
deviation occurs at a point at which, when the truncation order
tends to infinity, the analytical stable and unstable manifolds
should join each other smoothly. Here we establish this fact more
definitely by using higher order approximations and a much larger
number of digits in the calculations.

The above calculation indicates that in the case $\epsilon=0$ the
domain of convergence of the normalizing transformation $\Phi$
along the $\xi'$ (and $\eta'$) axis is finite. However, we can
provide a further indication by examining the domain of {\it absolute}
convergence, using, again, the D'Alembert sequence of radii
$\rho_n$, where, in the series yielding the transformation $\Phi$,
we set the book-keeping constant equal to $\lambda=1$ and we
rearrange all terms according to the polynomial orders $n=1,2,\ldots$
in the variables $(\xi,\eta)$. The evolution of the sequence
$\rho_n$ with $n$ is shown in Fig.\ref{mansep}b. We now see that,
contrary to the mapping case, in the present case the D'Alembert
sequence stabilizes at a finite constant value. Up to the
order $n=200$, we find that variations of the value of $\rho_n$
occur only in the fifth digit and beyond. Thus, we estimate the
D'Alembert radius as $\rho=1.1392...$. We can also obtain
an estimate of the domain of absolute convergence not only
along the axes $\xi'$ or $\eta'$, but also along any line of
the form $\eta'=\tan(\theta)\xi'$ passing from the origin,
for various angles $\theta$. In this case, setting $\xi'=s\cos\theta$,
$\eta'=s\sin\theta$, the transformation (\ref{trhyp}) for the
variable, say, $\xi$, for $\phi'=0$, leads to
\begin{equation}\label{xis}
\xi=g_1(\theta)'' s + g_2(\theta)'' s^2 + ...
\end{equation}
where $s$ is a length parameter along the chosen line. Thus,
implementing Eq.(\ref{rhodal}), but for the coefficients $g_i''$
which depend on $\theta$, we can estimate the D'Alembert radius
$\rho(\theta)$ in the particular direction. We have checked that
the D'Alembert radii $\rho_n$ along all directions have stabilized
up to several significant digits at the order $n=200$.
We then use the value $\rho_{200}$ as an estimate of the asymptotic
value $\rho$. Figure \ref{mansep}c shows the result, by computing
$\rho(\theta)$ for 360 values of $\theta$ from 1 to 360 degrees.
Plotting the points $(\xi',\eta')=(\rho\cos\theta,\rho\sin\theta)$,
we find the form of the boundary of the domain of absolute
convergence. This has a star-like structure, and we see that
it is finite for all angles $\theta$.

The fact that the absolute convergence of the hyperbolic
normal form series takes place in a finite domain in the
particular case of one degree of freedom systems like the pendulum
was proven by Johnson and Tucker (2011) using majorant series (see 
also their figure 2). Since such systems have no chaos, it must be 
stressed that the finite extent of the domain of convergence is a 
property of the series and it is not related to chaos.

Passing now to the case $\epsilon=1$, Figure \ref{eps1man}a shows
a computation of the invariant manifolds as above, but for the
truncation orders $r=20$, $r=50$, and $r=80$. In this case,
the number of terms in the corresponding series is so large that
it makes higher order calculations quite difficult at precisions
higher than the usual double precision level. In fact, we used a
fortran program in order to extend the computation to the 80-th
order using double precision. However, the accumulation of
round-off errors in the series terms results in a loss of accuracy
for some coefficients even at the first significant digit after
an order $r\sim 40$. This we were able to check by computing
the numerical differences between coefficients for which it is
known that, by the symmetry of the Hamiltonian (\ref{hrpr2})
with respect to the axis $\psi=0$, their values should be equal.

Returning to Fig.\ref{eps1man}a, the qualitative form of the
analytical manifolds is quite similar to the corresponding form in
the separatrix case (Fig.\ref{mansep}a). Namely, the analytical
manifolds clearly deviate from the true ones before reaching the
position of the homoclinic point $H_0$. This indicates that the
domain of convergence of the normalizing transformation along the
axes $\xi'$ and $\eta'$ is finite in this case as well. In fact, a
simple argument can show that the domain of absolute convergence,
for $\epsilon$ small, is close to the corresponding domain for
$\epsilon=0$, for all directions $\theta$. This is because we can
readily see that the corresponding series for, say, the variable
$\xi$, introduce only $O(\epsilon)$ corrections with respect to
the series (\ref{xis}), namely the new series are of the form
\begin{equation}\label{xiseps}
\xi=h_0(\theta)+(g_1(\theta)''+h_1(\theta)) s
+ (g_2(\theta)''+h_2(\theta)) s^2 + ...
\end{equation}
where all coefficients $h_i$ are of order $\epsilon$. Then, it is
trivial to prove that for $\epsilon$ sufficiently small the
D'Alembert sequence converges to a value $\rho+O(\epsilon)$, where
$\rho$ is the convergence radius of the original sequence. As
shown in Fig.\ref{eps1man}b, for $\epsilon=1$ the sequence
$\rho_r$ exhibits a clear tendency to convergence, although with
larger fluctuations up to the order 40 than the corresponding
sequence of the $\epsilon=0$ case (Fig.\ref{mansep}b). In this
case, since the sequence $\rho_n$ seems to stabilize tending
asymptotically from above to a constant value, we used a somewhat
smaller value than $\rho_{40}$, namely $\rho=0.95\rho_{40}$, as an
estimate of the radius of absolute convergence.

For $\theta=0$, we find the value $\xi'=\rho=2.4$, which
represents the radius of convergence in the direction along the
unstable manifolds. In the next section, we will present an
extended method that allows to parameterize the invariant
manifolds at radii beyond $\rho$. Using that method, we find that
the value $\xi'=\rho=2.4$ corresponds to the point A along the
invariant manifolds in Fig.\ref{eps1man}a, given by
$(\psi_A,p_{\psi,A})=(-0.6146,0.5208)$. We observe that the
analytic computation of the invariant manifolds starts deviating
from the numerical one a little before the point $A$. This
verifies the fact that the homoclinic point $H_0$ is outside the
domain of convergence. In fact, computing the hyperbolic series
for different values of $\epsilon$ yields that the distance of $A$
from $H_0$ increases as $\epsilon$ increases.

Returning to the computation of $\rho(\theta)$,
Fig.\ref{eps1man}c shows the corresponding plot for various angles
$\theta$ from 0$^{\circ}$ to 360$^{\circ}$. This is qualitatively
similar to Fig.(\ref{mansep}c), indicating that no important
differences regarding the convergence of the hyperbolic normal
form exist between the cases $\epsilon=0$ and $\epsilon=1$.

In conclusion, our numerical results indicate that, contrary to
the case of simple 2D mappings analytic over the whole plane, in
the Hamiltonian case the domain of convergence of the hyperbolic
normal form has a {\it finite} extent along the invariant
manifolds represented by the axes $\xi'=0$ and $\eta'=0$, even if
the original Hamiltonian is analytic over its entire phase space.
This appears at first to pose a severe limit to the usefulness of
the hyperbolic normal forms in this case. In fact, although the
calculations of Vieira and Ozorio de Almeida (1996)
seemed to indicate that it was possible to compute homoclinic
orbits using only the hyperbolic normal form series, a careful
inspection of their computations shows that the accurate location
of the homoclinic orbits was only possible by using the so-called
(by them) `z-propagation', which means numerical propagation of
initial conditions obtained in a small domain around the origin.
Thus, this computation was not based exclusively on the use of
series.

On the other hand, Ozorio de Almeida and Vieira (1997)
discussed also from a theoretical point of view the issue of the
extension of Moser's domain of convergence in the Hamiltonian
case. Their main argument uses the concept of analytic
continuation which allows to represent the flow in time of some
initial condition using a series polynomial in the canonical
variables. In the following section we will show that in the case
of Hamiltonian expressed in power series there are limits to the
domain of convergence. This explains the difference between simple
mappings and Hamiltonian systems. However, by modifying the method
of Ozorio de Almeida and Vieira (1997), we will provide
an extended method allowing the computation of the invariant
manifolds over a large extent in Hamiltonian models using only
series.

%%%%%%%%%%%%%%%%%%%%%%%%%%%%%%%%%%%%%%%%%%%%%%%%%%%%%%%%%%%
\section{Extended method}
%%%%%%%%%%%%%%%%%%%%%%%%%%%%%%%%%%%%%%%%%%%%%%%%%%%%%%%%%%%
Consider a Hamiltonian function $H(\psi,\phi,p,I)$ periodic
in $\psi,\phi$, and denote
by $q\equiv(\psi,\phi,p,I)$ a set of initial conditions
and by $q_t\equiv(\psi_t,\phi_t,p_t,I_t)$ their image
at a fixed time $t$ under the flow of $H$. Fixing $t$
we will consider the domain of analyticity of the mapping
in time of the phase-space variables, i.e. the mapping
$q_t=F_t(q)$. To this end, we express first $F_t(q)$ via
its Lie series expansion
\begin{equation}\label{ftlie}
q_t=\exp(L_{tH})q
\end{equation}
For all variables $(\psi_t,\phi_t,p_t,I_t)$ this takes the
form of a Fourier series. For example, we have
\begin{equation}\label{pt}
p_t=\exp(L_{tH})p =
\sum_{k_1,k_2} \tilde{f}_{p,(k_1,k_2)}(p,I,t)
e^{i(k_1\psi+k_2\phi)}
\end{equation}
with $\tilde{f}_{p,(k_1,k_2)}(p,I,0)$ equal to $p$, for $k_1=
k_2=0$, and zero otherwise (since the series (\ref{pt}) is
a near identity transformation). The Fourier
coefficients $\tilde{f}_{p,(k_1,k_2)}(p,I,t)$ are power
series of the time variable, i.e. we have
\begin{equation}\label{fpt}
\tilde{f}_{p,(k_1,k_2)}(p,I,t)=
\tilde{f}_{p,(k_1,k_2)}^{(0)}(p,I)+
t\tilde{f}_{p,(k_1,k_2)}^{(1)}(p,I)+
t^2\tilde{f}_{p,(k_1,k_2)}^{(2)}(p,I)+...
\end{equation}
For a fixed value of $t$, the domain of analyticity of
(\ref{pt}) intersects the action space at all points
$(p,I)$ for which:\\

i) The Taylor series (\ref{fpt}) is convergent, and\\

ii) There are positive constants $A(p,I),\sigma(p,I)$ such
that for all $k_1,k_2$ we have
$$
|\tilde{f}_{p,(k_1,k_2)}(p,I,t)|<Ae^{-(|k_1|+|k_2|)\sigma}.
$$

The last condition stems from the requirement that the
Fourier coefficients $\tilde{f}_{p,(k_1,k_2)}(p,I,t)$ decay
exponentially (see Giorgilli 2001, pp.88-90).

In a similar way we can determine the domains of analyticity
of the Lie mappings of the form (\ref{pt}) for all other
variables. The final domain of analyticity of the mapping
$F_t$ is given by the common intersection of all partial
domains. Let us denote this domain by $D_t$.

Assuming that the Hamiltonian $H$ is polynomial in the action
variables $(p,I)$, a key remark is now the following. For fixed
$(p,I)$, regarding the angles the Fourier series (\ref{pt})
are analytic in complex strips of the form $0\leq Re(\psi)<2\pi$,
$|Im(\psi)|<\sigma(p,I)$. This implies that if we Taylor-expand
(\ref{pt}) with respect to the angle $\psi$ corresponding to the
{\it hyperbolic degree of freedom} (cf. section 3, Eq.(\ref{hamex4}))
around some value $\psi_0\in\mathbf{T}$, the expansion converges
only in the interval $|\psi-\psi_0|<\sigma$. Therefore,  the
polynomial expansion of the mapping $F_t$ in the angular variable
corresponding to the hyperbolic degree of freedom  converges only
in a domain ${\cal D}_t$ which is a restriction of $D_t$, defined
by the above inequality.

We may finally consider the composition
\begin{equation}\label{FT}
F_{T,S_T}=F_{t_n}\circ F_{t_{n-1}}\circ\ldots\circ F_{t_1}
,~~~~~
S_T=\left\{t_1,t_2,\ldots,t_n:  \sum_{i=1}^n t_i=T\right\}
\end{equation}
which maps the variables $q$ to their values $q_T$ at the time
$t=T$ (the period of the stroboscopic Poincar\'{e} section) through
$n$ intermediate mappings at the times $t_1$, $t_1+t_2$,$\ldots$
$t_1+\ldots+t_n=T$. The domain of analyticity $D_{T,S_T}$ of the
mapping $F_{T,S_T}$ is a restriction of the domains of analyticity
of the mappings $F_{t_i}$, and its polynomial expansion in the angle
corresponding to the hyperbolic degree of freedom converges in a
restriction ${\cal D}_{T,S_T}$ of the domains ${\cal D}_{t_i}$.

We denote now by ${\Delta}_T$ the union of the sets ${\cal
D}_{T,S_T}$ for all possible choices of $S_T$ such that the sets
${\cal D}_{T,S_T}$ include the unstable periodic orbit $P$. Then,
according to Da Silva Ritter et al. (1987), the
normalizing transformation $\Phi$ of the hyperbolic normal form
around P is convergent in a domain of the new canonical variables
$(\psi',\phi',p',I')$ whose image is the set ${\Delta}_T$.
Following the argument of Ozorio de Almeida and Vieira
(1997), we have then the result that the domain of
convergence  of the hyperbolic normal form can be extended, from
the original Moser's domain, all over the domain ${\Delta}_T$.
However, as we have seen in the previous section, the numerical
indications are that the {domain ${\Delta}_T$ contains no
homoclinic points}. Thus, with the usual polynomial series
expansion for the angle $\psi$ we cannot go beyond such
homoclinic points.

If, however, {\it we maintain the Fourier form} of the transformations
like (\ref{pt}), without expanding the trigonometric terms with respect
to the angle $\psi$, then we can obtain a computation of the invariant
manifolds in a domain beyond ${\Delta}_T$. To this end, we define the
set of extended transformations:
\begin{equation}\label{compo}
\Phi_m = F_T^{m}\circ\Phi\circ N_T^{-m},
~~~m=0,\pm 1,\pm 2,\ldots
\end{equation}
where

i) $N_T$ is the mapping of the new canonical variables
$(\xi',\eta')$ after $m$ periods under the normal form
(\ref{nfhyp}). This is trivially found by the normal
form equations of motion (see section 3).

ii) we analyze $F_T$ as a composition of the form
(\ref{FT}), keeping each of the functions $F_{t_i}$
in its original form, given by a Fourier series of the
form (\ref{pt}).

In words, Eq.(\ref{compo}) means the following procedure,
which uses only series computations: starting from
a point $(\psi',\phi',p',I')$ in the new variables,
i) compute its m-th pre-image (or image, for $m<0$)) with
period $T$ under the normal form series dynamics, ii) use
the transformation series $\Phi$ to compute the values of
the old canonical variables corresponding to the last
point found in step (i), and finally iii) use the truncated
Fourier series $F_T^m$ in order to map forward (or backward,
for $m<0$) for $m$ periods the point found in (ii).

In the numerical examples given below, we choose $S_T$
as $S_T=\{T/4,T/4,T/4,T/4\}$. Other choices of the
separation of $T$ in a small number of parts are possible,
but we found that the choice $S_T=\{T/4,T/4,T/4,T/4\}$ is simple
and works with good precision for all initial conditions in the
whole domain where the asymptotic invariant manifolds extend.
Thus, by this choice we are able to compute the asymptotic
invariant manifolds over a large length using only series
(a preliminary computation of this type was done in 
Efthymiopoulos (2012b)).

%-------------------------------------------------------------
\begin{figure}
\begin{center}
\includegraphics[scale=0.7]{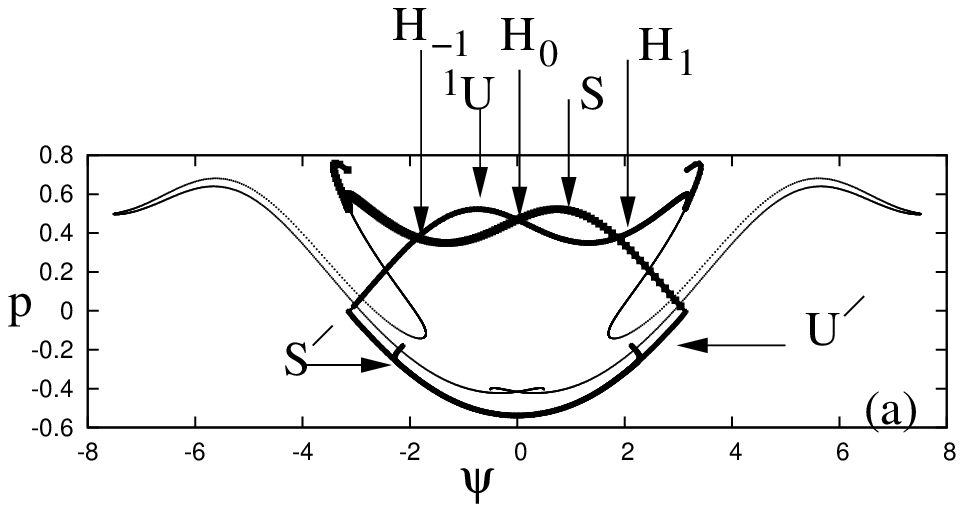}
\includegraphics[scale=0.7]{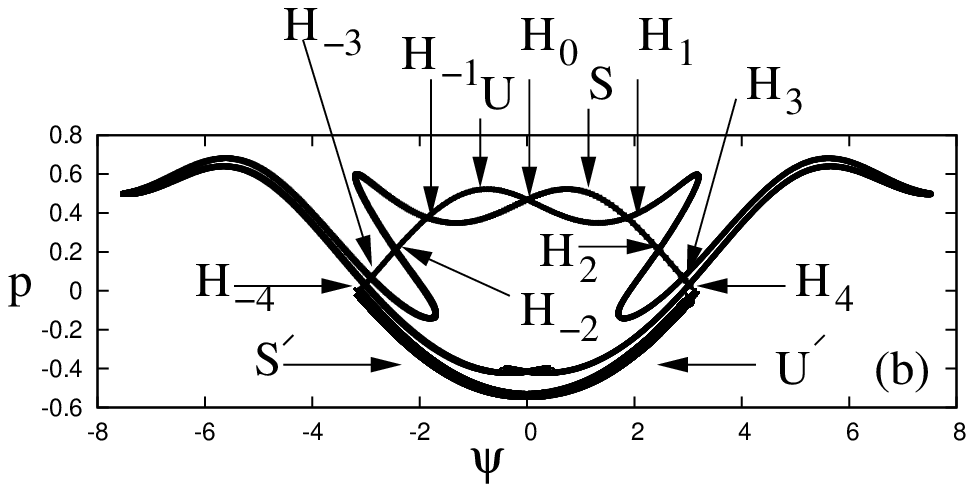}
\caption{Analytical invariant manifolds for $\epsilon=1$ computed
with the extended method setting (a) $m=1$ and (b) $m=3$. In the
first case, the analytical manifolds intersect at the first
and second homoclinic points $H_0$, $H_1$ (or $H_{-1}$), and
deviate afterwards. The position of these points can be computed
with high accuracy using only series. In the second case, the
analytical manifolds reproduce many features of the homoclinic
tangle, including the oscillations of at least third order lobes.}
\label{cont13man}
\end{center}
\end{figure}
%-------------------------------------------------------------
As an example, Figure \ref{cont13man} shows a computation of the
asymptotic manifolds when $\epsilon=1$, and $m$ in the
extended transformation (\ref{compo}) is chosen as $m=1$
(Fig.\ref{cont13man}a), or $m=3$ (Fig.\ref{cont13man}b) for the
unstable manifolds, and $m=-1$, $m=-3$ for the stable manifolds.
We now see that the analytical computation in the second case is
able to yield the homoclinic points $H, H_1, H_{-1}, H_2, H_{-2}...$,
corresponding to the intersections of the upper branches of
the invariant manifolds. In the same way, we can locate the
homoclinic intersections of the lower branches of the manifolds.
The latter, however, develop oscillations of a very small
extent, while the upper manifolds develop lobes defined by
oscillations, back and forth, over a scale about three times
as large as the whole extent of the separatrix of the unperturbed
case.

Since in practice the extended mappings (\ref{compo}) are computed 
by finite truncations of the Lie series for the mapping  $N_T^{-m}$ 
as well as the normalizing transformation $\Phi$, the accuracy of 
the extended method depends only on the order of truncation. 
We check the accuracy of $N_T$ by computing how well the energy 
is preserved in the mapping, under $N_T$, of initial conditions 
and their images on the stroboscopic surface of section. As 
an indication, at the 30th order of truncation the energy is 
preserved to about six digits close to the neighborhood of 
the first homoclinic point. This sets the overall precision 
of the method. 

The union of all extended mappings (\ref{compo}) for
$m=\pm 1,\pm 2,\ldots$ provides the complete parametrization
of the unstable or stable manifolds in terms of the `length'
parameters $\xi'$ or $\eta'$. In practice, however, we can only
choose finite values of $m$. Then, the parametrization
can be computed up to a finite maximum value of $\xi'$
(or $\eta'$), satisfying the relation $|\xi'|_{max} =
\rho \lambda^m$, where $\lambda=e^\nu T$, and $\rho$ is
the radius of convergence of the initial series of the
canonical transformation $\Phi$. In our model example,
we find $\lambda=5.91$, while (from Fig.\ref{eps1man}b)
$\rho\simeq 2.4$. Thus, with $m$ as high as $m=3$, we find
$|\xi|_{max}\simeq 5\times 10^2$, i.e. the parametrization
extends to a length about 200 times larger than the length
found by the original Moser normal form series. But if we take
$m$ even larger, we can find theoretically the asymptotic
curves up to an arbitrarily large extent.

These properties of the extended method have immediate
application in locating the homoclinic points and a number
of other features of the homoclinic tangle and its neighborhood.
In order to compute a homoclinic point, we consider the quantities
$\xi$, along the unstable manifold, and $\eta$, along the stable
manifold, as parameters whose values can be varied. We then use
a root-finding method (like Newton-Raphson) to compute
a homoclinic point as follows. Determining, via the extended
transformations (\ref{compo}) the transformations
$p(\xi,\eta),\psi(\xi,\eta)$, we solve simultaneously the
equations
\begin{equation}\label{homeq}
K(\xi,\eta)=p(\xi,0)-p(0,\eta)=0,~~~
L(\xi,\eta)=\psi(\xi,0)-\psi(0,\eta)=0~~.
\end{equation}
In the use of Newton-Raphson, all partial derivatives appearing in
the extended transformation are given in the form of truncated
series, since, by the chain rule, the Jacobian matrix
$\partial(K,L)/\partial(\xi,\eta)$ can be computed by the product
of the Jacobian matrices of all functions participating in the
function composition (\ref{compo}). In particular, expressing 
the part $N_T^{-m}$ as Lie series is crucial in analytically 
computing the parametrizations used in the quantities $K,L$ 
as well as their derivatives. As a result, the accuracy of the
present method is only limited by the accuracy of the various
truncations in the computed series. For example, we compute the
position of the homoclinic point $H_0$ in Fig.(\ref{eps1man}a)
using the extended transformation (\ref{compo}) for $m=1$. We have
$\psi=0$, while for $p_{H_0}$ we find $p=0.5592605371492...$ with
a precision of 13 significant digits.

%-------------------------------------------------------------
\begin{figure}[t]
\begin{center}
\includegraphics[scale=0.8]{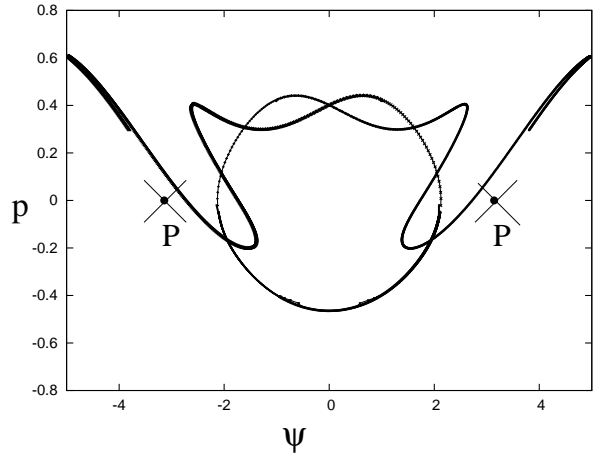}
\caption{A curve of the form $\xi'\eta'=c$, mapped in the old 
variables, for $c=-0.15$, computed with the extended transformation 
$m=2$. The unstable periodic orbit is represented by the point P.}
\label{cont1cv}
\end{center}
\end{figure}
%-------------------------------------------------------------
Finally, the extended mappings (\ref{compo}) offer the possibility
to compute curves in the chaotic domain corresponding to values of
$c$ different from zero. Such curves can be `invariant' or 
`quasi-invariant'. Namely, while the hyperbolic normal form converges 
all along a hyperbola in the variables $\xi',\eta'$ defined by 
$\xi'\eta'=c$, with $c\neq 0$, it is possible that only finite 
segments of the hyperbola map onto segments of the same
hyperbola within the domain of convergence of the normalizing 
transformation $\Phi$. This implies that, even using the extended 
transformations $\ref{compo}$, one may not be able to define a 
continuous set that maps exclusively onto itself at successive 
iterations. On the other hand, for small values of $c$ the 
curves $\xi'\eta'=c$ are invariant. In practice, it is difficult 
to find the limiting value of $c$ up to which we obtain such 
invariant curves. 

To our knowledge, the properties of such curves have not so far been 
studied. Figure \ref{cont1cv} shows an example of the form of the 
curves corresponding to the value $c=\xi'\eta'=-0.15$ 
in the model (\ref{hrpr2}) (we find that negatives values of $c$ 
correspond to curves in the interior domain defined 
by the invariant manifolds up to the  homoclinic intersection $H_0$,
while $c>0$ corresponds to curves in the exterior domain). In
order to compute the curves, we use a unique parameter, say,
$s=\xi$, $\eta=c/s$, and make sure that the mapping $N_T^{m}$ 
(for $m$ positive or negative) brings the points corresponding to 
the chosen range of values of $s$ within the domain of convergence 
of Moser's series for the transformation $\Phi$.

In Fig.\ref{cont1cv} we observe that the curve of $c=-0.15$
follows in general a path parallel to the paths of the asymptotic
manifolds (cf. Fig.\ref{cont13man}). Furthermore, the curves with 
$c\neq 0$ have a number of self-intersections. Such intersections 
define points accumulating to some homoclinic points as 
$c\rightarrow 0$. However, there are also intersections developed 
between the curves, that do not accumulate around homoclinic points. 
In fact, similarly to Da Silva Ritter et al. (1987), we can use
the computation of the intersections between one or more
curves, for $c\neq 0$, in order to compute high order periodic 
orbits accumulating to some homoclinic points. The study of such 
orbits is of particular interest, since they are related to the 
definition and statistics of the {\it recurrence times} in the 
domain of homoclinic chaos (see Contopoulos and Polymilis 1993). 
Their study using analytical series methods is proposed for 
future work.

\section{Unstable equilibria in a polynomial Hamiltonian model}

The example of the Hamiltonian (\ref{hrpr2}) dealt with in
sections 3 and 4 represents a time-dependent Hamiltonian system
which is reduced to an autonomous two-degrees of freedom system
via the use of the action-angle variables $(I,\phi)$.
Nevertheless, this system differs from the systems considered in
the theorems of Moser (1958) and Giorgilli (2001) because, if we
introduce the Cartesian canonical variables $x=\sqrt{2I}\sin\phi$,
$p_x=\sqrt{2I}\cos\phi$, the Hamiltonian (\ref{hrpr2}) is not
polynomial in $(x,p_x)$. Then, one may wonder whether the results
of sections 3 and 4 are applicable in the polynomial case
considered by Moser and Giorgilli.

%-------------------------------------------------------------
\begin{figure}[t]
\begin{center}
\includegraphics[scale=0.8]{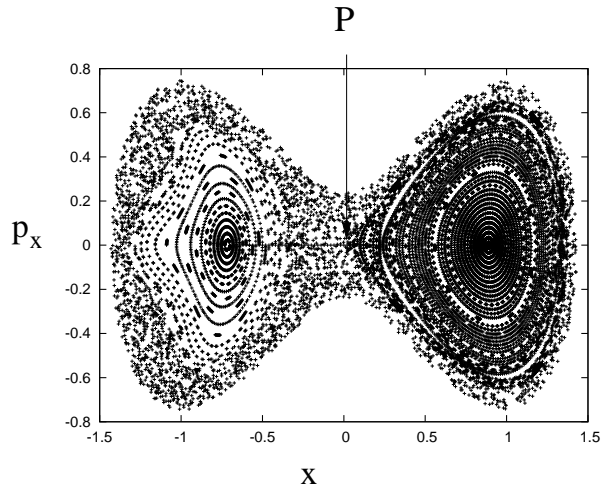}
\caption{The Poincar\'{e} surface of section $(x,p_x)$, for $y=0$,
$p_y>0$, in the Hamiltonian model (\ref{hampol}), for the energy
$E=H=0.03$. The main unstable periodic orbit is represented by the
point P.} \label{polsect}
\end{center}
\end{figure}
%-------------------------------------------------------------
In order to investigate this question, we consider the Hamiltonian
model
\begin{equation}\label{hampol}
H={1\over 2}(p_x^2-x^2)+{1\over 2}(p_y^2+y^2) +x y^2 + x^2 y^2 +
x^4/4~~~.
\end{equation}
The Hamiltonian flow of (\ref{hampol}) yields an unstable
equilibrium at $x=y=p_x=p_y=0$. The corresponding energy is
$E=H=0$. On the other hand, for $E>0$ the unstable equilibrium
solution is continued as an unstable periodic orbit solution.
Figure \ref{polsect} shows a Poincar\'{e} surface of section
$y=0$, $p_y>0$ at the energy $E=0.03$. The periodic orbit (denoted
by P) intersects the Poincar\'{e} section at $x_P=0.0246122$,
$p_{x,P}=0$. We observe the formation of a figure-eight type
chaotic layer due to the homoclinic chaos around P.

We now implement the algorithm described in subsection 3.2 in
order to compute analytically the invariant manifolds emanating
from P. In this case, the algorithm becomes equivalent to the
algorithm described in Giorgilli (2001) after the substitution
$y=(2I)^{1/2}\sin\phi$, $p_y=(2I)^{1/2}\cos\phi$,
$x=(\xi-\eta)/\sqrt{2}$, $p_x=(\xi+\eta)/\sqrt{2}$, $\nu=1$.
Finally, we choose our book-keeping to follow the ordering by
polynomial degree, i.e., terms of book-keeping order $s$
correspond to terms of degree $s+2$ in the variables
$(y,p_y,\xi,\eta)$.

We compute the hyperbolic normal form (Eq.(\ref{nfhyp})), as well
as the corresponding near-identity transformations
(Eq.(\ref{trhyp})) up to the order $s=43$. In the new canonical
variables $(\xi',\eta',\phi',I')$ the first few terms of the
normal form read
\begin{equation}\label{nfpol}
E=Z_h=I'+\xi'\eta' - 0.6 I'\xi'\eta'+ 0.55 I'^2 ...
\end{equation}
We note that no small parameter (like $\epsilon$ in the model of
sections 3 and 4) appears in the normal form (\ref{nfpol}). Here,
however, the periodic orbit P, and its associated manifolds, are
parameterized by the value of the energy $E$, or of the action
$I'=I_0$ found by setting $\xi'=\eta'=0$ in Eq.(\ref{nfpol}) and
solving the equation for $I'$ for a given value of $E$. For
$E=0.03$ we find $I_0=0.0295075$. The corresponding period is
given by $T_0=2\pi/\omega(I_0)=6.07748$, where $\omega =
dZ_h/dI_0$.

In the computation of the invariant manifolds, it is convenient to
determine as surface of section one in which all initial
conditions on the manifolds return at equal times $t=T_0$. This
surface of section corresponds to the choice $\phi'=0$, for points
of the unstable manifold, or $\phi'=\pi$ for points of the stable
manifold.

Setting $I'=I_0,\phi'=0$, we obtain all four canonical variables
$x,p_x,y,p_y$ as polynomial series of the two variables
$\xi'$,$\eta'$. An important consistency check is to replace these
series in the Hamiltonian (\ref{hampol}). Then, we find that all
series coefficients cancel (to a numerical precision $10^{-15}$,
leaving only a constant equal to 0.03, i.e., the numerical value
of the energy corresponding to $I_0$.

%-------------------------------------------------------------
\begin{figure}
\begin{center}
\includegraphics[scale=0.7]{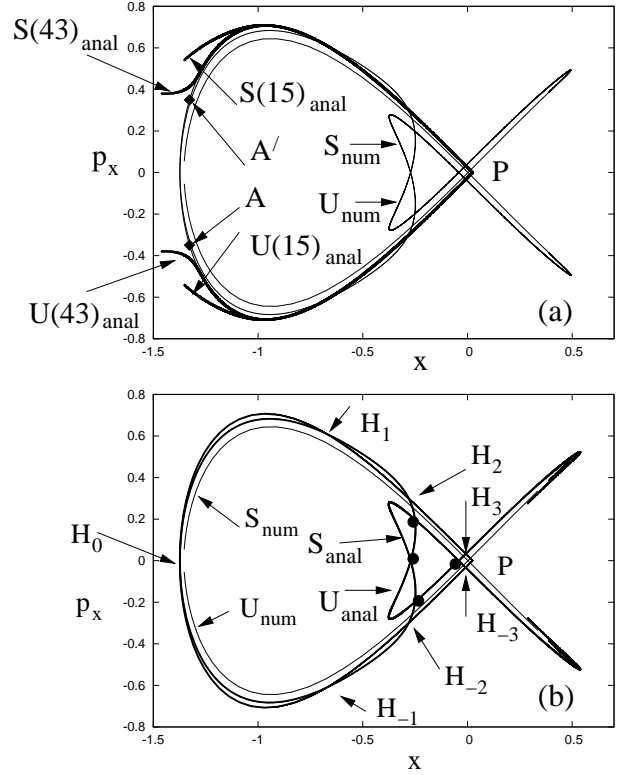}
\caption{(a) Analytical (thick lines) versus numerical (thin
lines) invariant manifolds in the model (\ref{hampol}), at the
energy $E=0.03$, in the surface of section given by $\phi'=0$, for
the unstable manifold, or $\phi'=\pi$ for the stable manifold,
where $\phi'$ is the angle in the new variables in the elliptic
degree of freedom after implementing the normal form
transformations (see text). The analytical manifolds are computed
at the truncation orders $r=15$ and $r=43$. The points A and A'
represent the limit of convergence of the series along the
unstable and stable manifolds respectively. (b) Same as in (a),
but now with for the analytical manifolds computed by the extended
method of section 4. The analytical manifolds now reproduce well
several lobes of the numerical ones, and they can be used to
locate the position of the homoclinic points $H_0$ and $H_{\pm
j}$, $j=1,2,3$. The thick dots represent higher order homoclinic
points reproduced by the same computation.} \label{hyppol}
\end{center}
\end{figure}
%-------------------------------------------------------------
Figure \ref{hyppol}a shows, now, the computation of the invariant
manifolds of P numerically, as compared with the analytic
manifolds produced by the hyperbolic normal form series truncated
at the orders 15 and 43. We observe that, as the order of
truncation increases, the analytic unstable manifold tends to
coincide with the numerical one up to a point that comes closer
and closer to the point A in Fig.\ref{hyppol}a. This point
corresponds to an approximation of the domain of convergence of
the normal form transformation series, as computed numerically
using, again, d'Alembert's criterion. Similarly, the analytic
stable manifold, which is symmetric with respect to the unstable
one with respect to the axis $p_x=0$, starts deviating from the
numerical one a little before A', i.e. the point symmetric to A.

Figure \ref{hyppol}b, now, shows the result of using the extended
method of section 4 in the present example. Here, we implemented
Cauchy's criterion in order to determine the distance from the
origin of the nearest singularity of the Hamiltonian flow in the
complex time domain for all initial conditions in the range
$-1.5\leq x\leq 1.5$, $-1\leq p_x\leq 1$. We find that all
singularities are at distances larger than $|t|=1$. Thus, we used
a partition in eight time steps of length $\Delta t = T_0/8$, to
ensure that the extended series converge for all initial
conditions considered. We observe that the extended series are
able to reproduce the numerical asymptotic curves up to a length
covering at least the homoclinic points $H_0$, $H_1$, $H_2$,
$H_3$, and $H_{-1}$, $H_{-2}$, $H_{-3}$. Furthermore, for the used
value of $E$, the analytical manifolds are also able to reproduce
higher order homoclinic intersections developed among some low
order lobes of the invariant manifolds, whose size, at this
particular energy, is sufficient for them to intersect.

In conclusion, the main conjectures and results found for generic
action-angle systems of the form considered in our present
formalism appear also in the original case examined by Moser,
i.e., of polynomial Hamiltonian systems with unstable equilibria.

%%%%%%%%%%%%%%%%%%%%%%%%%%%%%%%%%%%%%%
\section{Conclusions and Discussion}
%%%%%%%%%%%%%%%%%%%%%%%%%%%%%%%%%%%%%%

We studied the convergence properties of the hyperbolic normal
form series used in the computation of the invariant manifolds of
an unstable equilibrium point, or an unstable periodic orbit, in
2D mappings and in 2D generic Hamiltonian systems  expressed in 
action-angle variables. Our main findings can be summarized as 
follows:

1) In 2D mappings analytic over their whole phase space,
we confirmed by numerical experiments the statement of Da Silva
Ritter et al. (1987) that the convergence of the normal
form extends to infinity along the invariant manifolds, and to
finite, but large domain, for nearby `quasi-invariant' curves.
This allows in practice for the computation of many interesting
features of the homoclinic tangle using only series (e.g.
homoclinic points, periodic orbits of high period, high order
lobes of the invariant manifolds etc.)

2) In two different 2D mappings that we studied, the rate by which
the sequence of D'Alembert radii of the formal series goes to
infinity is found to obey different laws. In the standard
map it is found to grow logarithmically with the normalization
order $r$, while in the H\'{e}non map it is quadratic in $r$.
However, when back-transforming to the original variables, in
both cases the length along the manifolds recovered by the
truncated analytical series increases with the truncation order
$r$ as $\log(r)$. Finally, if a mapping has a limited analyticity
domain, then the sequence of D'Alembert radii tends to a constant,
indicating that the domain of convergence in this case is finite.

3) As a first example of the Hamiltonian case, we studied a
pendulum model with periodic perturbation, which cannot be
reduced to a polynomial form. We find a finite domain of
convergence of the hyperbolic normal form, and we provide
results indicating that this  domain does not reach any
homoclinic point. We also give the form of the domain of
convergence on the plane of the new hyperbolic canonical
variables. Finally, we checked the accuracy of our computations
when computing the normal form series using the usual double
precision, or 80-digit precision. The latter is needed in order to
control the propagation of round-offs in some cases, beyond a high
normalization order.

4) Then, we proposed a new method by which one can compute
extensions of the original normalizing transformations of Moser
using only series. Thus we materialize the idea of analytic
continuation proposed by Ozorio de Almeida and Viera (1997). 
However, contrary to these authors, we choose action angle variables 
to represent the series yielding a part of the extended transformation 
(namely the mappings $F^m$ in Eq.(\ref{compo})), without expanding 
the trigonometric functions of the angle corresponding to the hyperbolic 
degree of freedom in power series. In this way, it becomes possible 
to parameterize the invariant manifolds up to an arbitrarily large
extent. We show the efficiency of the method by concrete numerical
examples.

5) We emphasized the fact that the possibility to extend the
normalizing transformations to a large domain allows one to study
features of chaos using only series. In particular, we approximate
the intricate lobes formed by the asymptotic manifolds near the
unstable point, and we compute curves of the form $c=\xi'\eta'$  
(mapped in the old variables), which are near the asymptotic 
manifolds. We call these features `the structure of chaos'. 
We propose to use the same approach in order to compute high order 
periodic orbits that accumulate around one or more homoclinic points.

6) We finally study the case of a polynomial Hamiltonian
model, in which the original normal form of Moser is applicable.
In this case as well we confirm the results of sections 3, and 4,
namely i) we provide numerical evidence that the convergence
domain of Moser's normalizing transformations does not reach any
homoclinic point, and ii) that an extension using our method of
section 4 allows to parameterize the manifolds over an extent long
enough to include several homoclinic points.

The results found so far point towards a number of applications
and extensions. In particular, we emphasize two benefits from
the extended transformations (\ref{compo}): i) the possibility
to obtain a parametrization of the invariant manifolds over an
arbitrary length, and ii) the possibility, in numerical implementations,
to obtain the manifolds with uniform linear sampling density over any
desired length, using higher values of the multiplicity $m$ (see
Eq.(\ref{compo})) to achieve the same density for a longer length.
Property (i) marks the essential difference between our new method
and previous numerical methods that simply propagate forward initial
conditions obtained by accurate calculations within the domain of
convergence of the original method of Moser. Property (ii), on the
other hand, is useful in practical computations and visualizations
of asymptotic orbits, up to an arbitrarily long extent.

On the other hand, the possibility to represent analytically not
only the asymptotic manifolds, but also other invariant manifolds
in the neighborhood of an unstable orbit poses a number of
interesting theoretical questions related to the structure of
chaos in such a neighborhood. In particular, it is known
(Contopoulos et al. 1996) that tiny islands of stability can exist
arbitrarily close to the unstable fixed point. Such islands are
found both outside and inside lobes. According to the Newhouse
theorem (Newhouse 1977) such islands are generated by new (irregular)
periodic orbits which appear near points of tangency between the
stable and unstable asymptotic manifolds. In systems with a
compact phase space, such points are generated continuously, for
arbitrarily high values of the non-linearity parameter (Contopoulos 
et al. 1994). In fact, the method of analytic computation of the 
invariant manifolds lends itself to the computation of such points
of tangency of an arbitrarily high order. However, the curves 
$\xi'\eta'=c$ for $c\neq 0$ cannot represent the islands of the 
stable irregular periodic orbits.

In the model studied in the present paper, the asymptotic
invariant manifolds emanating from the periodic orbit at
$\psi=-\pi$ can be formally considered as forming heteroclinic
intersections with the manifolds emanating from the periodic orbit
at $\psi=\pi$. However the second orbit is the same with the first
one modulo $2\pi$. On the other hand, the present method can be
used with little modification in order to study true heteroclinic
connections between more than one periodic orbits. In fact, this
is the case in the model studied by Vieira and Ozorio de Almeida (1996),
which possesses a triple periodic orbit. Similar results can be
found for unstable periodic orbits of any multiplicity.

There are also examples (e.g. Polymilis et al. 2003) in which, after a
period doubling bifurcation, the homoclinic connections between
some periodic orbit evolve, by varying one parameter, to
heteroclinic connections between two different unstable periodic
orbits. The transition from homoclinic to heteroclinic dynamics
takes place at the point of the period doubling. The study of such
connections by analytical invariant manifolds can provide
important information on the kinds and statistics of the
recurrences, as well as the transport phenomena taking place in
the heteroclinic tangle and its neighborhood.\\
\\
{\bf Acknowledgements:} This research has been supported in part by 
the Research Committee of the Academy of Athens (grant 200/815).

{}


\begin{thebibliography}{}

\bibitem{Bart78}
Bartlett, J.H.: 1978, Instability of an area-Preserving Polynomial Mapping, 
Cel. Mech. 17, 3.

\bibitem{Bart82}
Bartlett, J.H.: 1982, Limits of stability for an area-preserving polynomial 
mapping, Cel. Mech. 28, 295.

\bibitem{Bart89}
Bartlett, J.H.: 1989, Almost stable regions for an area-preserving mapping, 
Cel. Mech. Dyn. Astron., 46, 129.

\bibitem{bazetal1993}
Bazzani, A., Giovannozzi, M., Servizi, G., Todesco, E., and
Turchetti, G.: 1993, Resonant normal forms, interpolating Hamiltonians and 
stability analysis of area preserving maps, Physica D, 64, 66.

\bibitem{beletal2010}
Bell\'{o}, M., G\'{o}mez, G., and Masdemont, J.J.: 2010, Invariant Manifolds, 
Lagrangian Trajectories and Space Mission Design in Perozzi, E., and 
Ferraz-Mello, S. (eds), Space Manifold Dynamics, Springer.

\bibitem{Bong01}
Bongini, L., Bazzani, A., and Turchetti, G.: 2001, Analysis of a model for 
resonant extraction of intense beams by normal forms and frequency map, 
Phys. Rev. Sp. Topics, 4, 114201.

\bibitem{bru1971}
Bruno, A.D., 1971, Trans. Moscow Math. Soc. 25, 131.

\bibitem{bru1989}
Bruno, A.D., 1989, Local Methods in Nonlinear Differential
Equations, Springer, Berlin.

\bibitem{cabetal2005}
Cabr\'{e}, X., Fontich, E., and de la Llave, R., 2005,The parameterization 
method for invariant manifolds III: overview and application, Jour. Diff.
Eq., 218, 444.

\bibitem{campetal2012}
Campagnola, S., Sherritt, P., and Russell, R.P, 2012, Flybys in the Planar Circular 
Restricted Three Body Problem, Cel. Mech. Dyn. Astron., 113, 343.

\bibitem{che1926}
Cherry, T.M., 1926, On the solutions of Hamiltonian systems in the neighborhood 
of a singular point, Proc. London Math. Soc. Ser.2, 27, 151-170.

\bibitem{conto90}
Contopoulos, G.: 1990, Asymptotic curves and escapes in Hamiltonian systems,
 Astron. Astrophys., 231, 41.

\bibitem{conto93}
Contopoulos, G., and Polymilis, C. 1993, Geometrical and dynamical properties 
of homoclinic tangles in a simple Hamiltonian system, Phys. Rev. E, 47, 1546.

\bibitem{conto94}
Contopoulos, G., Papadaki, H., and Polymilis, C. 1994,	
The structure of chaos in a potential without escapes,
Cel. Mech. Dyn. Astron. 60, 249.

\bibitem{conto96}
Contopoulos, G., Grousouzakou, E., and Polymilis, C. 1996,
	Distribution of Periodic Orbits and the Homoclinic Tangle,
Cel. Mech. Dyn. Astron. 64, 363.

\bibitem{conto02}
Contopoulos, G. 2002, Order and Chaos in Dynamical Astronomy,
Springer, Berlin.

\bibitem{ritter87}
Da Silva Ritter, G.I., Ozorio de Almeida, A.M., \& Douady, R.,1987,  
Analytical determination of unstable periodic orbits in area preserving maps
Physica D, 29, 181.

\bibitem{dellaz2005}
Delshams, A., and L\'{a}zaro, J.T., 2005, Pseudo-normal form near
saddle-center  or saddle-focus equilibria, J. Differential
Equations, 208, 312.

\bibitem{dephen1969}
Deprit, A., and Henrard, J., 1969, Construction of Orbits Asymptotic to a 
Periodic Orbit, Astron. J., 74, 308.

\bibitem{efthy12a}
Efthymiopoulos, C., 2012a, Canonical perturbation theory, stability and 
diffusion in Hamiltonian systems: applications in dynamical astronomy, in
P. M. Cincotta, C. M. Giordano and C. Efthymiopoulos (eds), Third La Plata 
Internat. School on Astron. Geophys.: Asociaci\'{o}n Argentina de Astronom\'{i}a.

\bibitem{efthy12b}
Efthymiopoulos, C.: 2012b, Hyperbolic Normal Forms and Invariant Manifolds. 
Astronomical Applications, Serbian Astr. J., 184, 1.

\bibitem{evans04}
Evans, T.E., Roeder, R.K.W., Carter, J.A., and Rapoport, B.I., 2004, 
Homoclinic tangles, bifurcations and edge stochasticity in diverted tokamaks,
Contrib. Plasma Phys., 44, 235.

\bibitem{franc81}
Franceschini, V., and Russo, L.: 1981, Stable and unstable manifolds of the 
Henon mapping, J. Stat. Phys., 25, 757.

\bibitem{gio01}
Giorgilli, A 2001, Unstable equilibria of Hamiltonian systems, Disc. Cont. Dyn. 
Sys., 7, 855.

\bibitem{gio02}
Giorgilli, A. 2002, Notes on exponential stability of Hamiltonian systems,
Pubblicazioni della Classe di Scienze, Scuola Normale Superiore, Pisa.
Centro di Ricerca Matematica "Ennio De Giorgi".

\bibitem{gomezetal2001}
G\'{o}mez, G., Jorba, A., Masdemont, J., and Sim\'{o}, C., 2011, 
Dynamics and Mission Design near Libration Points, Vol.IV, Advanced 
Methods for Triangular Points, Ch.3, World Scientific. 

\bibitem{gomez11}
G\'{o}mez, G., and Barrab\'{e}s, E. 2011, Space Manifold dynamics, 
Scholarpedia 6(2), 10597.

\bibitem{groross2009}
Grover, P., and Ross, S. 2009, Designing Trajectories in a Planet-Moon 
Environment using the Controlled Keplerian Map, Journal of Guidance, 
Contol and Dynamics 32, 436.

\bibitem{johnson11}
Johnson, T., and Tucker, W., Qual.: 2011, Theory Dyn. Sys. 10, 107.

\bibitem{jormas1999}
Jorba, \`{A}., and Masdemont, J., 1999,Dynamics in the center manifold of the 
collinear points of the restricted three body problem, Physica D, 132, 189.

\bibitem{moser56}
Moser, J 1956, Commun. Pure Applied Math., 9, 673.

\bibitem{moser58}
Moser, J 1958, Commun. Pure Applied Math., 11, 257.

\bibitem{new77}
Newhouse, S.E.: 1977, Am. J. Math., 99, 1061.

\bibitem{almeida97}
Ozorio de Almeida, A.M., and Vieira, W.M. 1997, Extended convergence of 
normal forms around unstable equilibria, Phys. Lett. A,
227, 298.

\bibitem{per10}
Perozzi, E., and Ferraz-Mello, S. 2010, Space Manifold
Dynamics, Springer.

\bibitem{poi1890}
Poincar\'{e}, H., 1890, Journal de Mathematiques Pures et
Apliqu\'{e}es, 6, 313.

\bibitem{pol03}
Polymilis, C., Contopoulos, G., and Dokoumetzidis, A., 2003,
The Homoclinic Tangle of A 1:2 Resonance in a 2-D Hamiltonian System,
Cel. Mech. Dyn. Astron., 85, 105.

\bibitem{roeder03}
Roeder, R.K.W., Rapoport, B.I., and Evans, T.E.: 2003,
Explicit calculations of homoclinic tangles in tokamaks,
Phys. Plasmas, 10, 3796.

\bibitem{rom90}
Rom-Kedar, V., 1990, Transport rates of a class of two-dimensional maps and flows, 
Physica D., 43, 229.

\bibitem{rossschee2007}
Ross, S.D., and Scheeres, D.J., 2007, Multiple Gravity Assists, Capture and Escape 
in the Restricted Three Body Problem, SIAM J. Appl. Dyn. Syst., 6(3), 576.

\bibitem{simo90}
Sim\'{o}, C.: 1990, On the Analytical and Numerical Approximation of Invariant 
Manifolds, in D. Benest and C. Froeschl\'{e} (eds),
`Les M\'{e}thodes Modernes de la Mac\'{a}nique C\'{e}leste,
Editions Frontieres, Paris.

\bibitem{vieira96}
Vieira, W.M., and Ozoiro de Almeida, A.M., 1996, Study of chaos in hamiltonian 
systems via convergent normal forms, Physica D, 90, 9.

\end{thebibliography}
\end{document}